\documentclass[12pt]{iopart}

\newcommand{\beq}{\begin{equation}}
\newcommand{\eeq}{\end{equation}}
\newcommand{\beqn}{\begin{eqnarray}}
\newcommand{\eeqn}{\end{eqnarray}}

\newcommand{\llabel}[1]{\label{#1}}              % DO NOT show equation label
\newcommand{\labeq}[2]{ \begin{equation} \llabel{#1}{#2}
\end{equation}}
\newcommand{\eqref}[1]{(\ref{#1})}

\usepackage{graphicx}
\usepackage{epsf}
\usepackage{longtable}
\usepackage{hyperref}
\usepackage{graphics,epsfig,placeins,subfigure,wrapfig}
\usepackage[usenames]{color}
\usepackage{amssymb}
\usepackage{soul}

\begin{document}
\title[GR simulations of compact binary mergers as sGRB
  engines]{General relativistic simulations of compact binary mergers
  as engines of short gamma-ray bursts}

\author{Vasileios Paschalidis$^*$} 
\address{Department of Physics, Princeton University, Princeton,
  NJ 08544, USA}
\ead{$^*$vp16@princeton.edu}

\begin{abstract}
Black hole - neutron star (BHNS) and neutron star - neutron star
(NSNS) binaries are among the favored candidates for the progenitors
of the black hole - disk systems that may be the engines powering
short-hard gamma ray bursts. After almost two decades of simulations
of binary NSNSs and BHNSs in full general relativity we are now
beginning to understand the ingredients that may be necessary for
these systems to launch incipient jets. Here, we review our current
understanding, and summarize the surprises and lessons learned from
state-of-the-art (magnetohydrodynamic) simulations in full general
relativity of BHNS and NSNS mergers as jet engines for short-hard
gamma-ray bursts.
\end{abstract}

\section{Introduction}

The LIGO and Virgo collaborations recently announced the detection of
two gravitational wave signals that were consistent with the inspiral
and merger of binary black hole
systems~\cite{LIGO_first_direct_GW,LIGOseconddirect}. A third signal,
also consistent with a binary black hole, was announced but was not
significant enough to be classified as a
detection~\cite{LIGOseconddirect}. These observations are milestones
in the field of gravitational physics because they confirmed for the
first time the validity of general relativity in the strong-field,
dynamical regime, they provided the cleanest evidence for the
existence of black holes and binary black holes, and gave us hints on
plausible formation scenarios for such systems. Most importantly,
these spectacular detections opened up a new window to observing our
Universe, and marked the onset of the era of gravitational wave (GW)
astronomy. Over the next few years, advanced LIGO is anticipated to
reach its design sensitivity and advanced VIRGO will join the
observations. As a result many more GW signals are anticipated to be
detected, and not only from other black hole binaries, but also from
the inspiral and merger of neutron star--neutron star (NSNS) and black
hole--neutron star (BHNS) binaries.
 
Coalescing NSNSs and BHNSs are not only sources of GWs, but also of
electromagnetic (EM) signals counterpart to the GWs that can arise
both before~\cite{Hansen:2000am,McWilliams:2011zi,Paschalidis:2013jsa,
  PalenzuelaLehner2013,2014PhRvD..90d4007P,Metzger2016MNRAS.461.4435M}
and after~\cite{MetzgerBerger2012,2015MNRAS.446.1115M} the GW peak
amplitude. Detecting both GW and EM signals that are generated from
the same source would provide a wealth of information about the
source, and allow novel tests of relativistic gravitation and
fundamental physics\footnote{Tests of relativistic gravitation with GW
  astronomy is an entire topic by itself, which will not touch upon in
  this review, but for some recent work and reviews see
  e.g.~\cite{PPE2009PhRvD..80l2003Y,Yunes2010PhRvD..82h2002Y,2011PhRvD..84f2003C,2012PhRvD..85h2003L,2012PhRvD..86h2001V,2012PhRvD..86b2004C,2013LRR....16....9Y,2013IJMPD..2241012A,2014PhRvD..90l4091S,ShibtaST2014PhRvD..89h4005S,2015PhRvD..91h4038P,Berti2015CQGra..32x3001B,Calabrese2016CQGra..33p5004C,Yunes2016arXiv160308955Y,2016arXiv160909825D,2016CQGra..33m5002G,2016PhRvL.116x1104B,2016PhRvD..93l4004S,2016PhRvD..93f4005R,2016CQGra..33e4001Y,2015arXiv151003845Y,2014IJMPD..2343009B,GW150914TestGR2016}
  and references therein.} GWs can also constrain the NS equation of
state (EOS) (see
e.g.~\cite{Lackey2015PhRvD..91d3002L,2016arXiv160703540B} for recent
work and references therein), and combination of GWs with EM signals
can help explain where r-process elements in the Universe may
form~\cite{Rosswog:1998gc}, and even allow for an accurate and
model-independent computation of the Hubble constant, and hence
constrain dark energy models~\cite{Nissanke2013arXiv1307.2638N}. In
fact, performing several of the aforementioned tasks may actually {\it
  require} an EM counterpart to the GW signal. For example, if gravity
behaves phenomenologically as in some scalar-tensor theory models,
where the deviations from general relativity (GR) may show only near
merger, degeneracies with the equation of state cannot be lifted by
GWs alone, whereas even partial information from EM counterparts can
lift the degeneracy~\cite{2015PhRvD..91h4038P}.

Detection of GW, EM and/or potentially even neutrino signals from the
same source would mark the onset of the era of ``multimessenger''
astronomy. However, the interpretation of multimessenger signals from
compact binary mergers will depend crucially on our theoretical
understanding of these events, which in turn requires simulations in
full relativistic gravitation to treat the strong, dynamical fields and
high velocities that naturally arise in these mergers.

Among all types of EM signatures NSNSs and BHNSs are thought to be
able to generate, a short gamma-ray burst (sGRB) is the one these
systems are best known. It has long been hypothesized that mergers of
these compact binaries can form the engine that may power an sGRB - an
accretion disk onto a spinning black hole
~\cite{EiLiPiSc,NaPaPi,MoHeIsMa,prs15,Meszaros:2006rc,LeeRamirezRuiz2007,Nakar2007PhR...442..166N}.
Advances in observations of sGRBs which led to the identification of
their host environments indicate that the progenitors of sGRBs are
mainly associated with elliptical galaxies and stem from an old,
evolved population of
stars~\cite{Fong2013ApJ...776...18F,Berger2014}. This makes the case
of NSNSs and BHNSs being the progenitors of sGRBs even more
compelling. Association of a sGRB with a GW signal consistent with the
inspiral and merger of a BHNS or NSNS system (perhaps the holy grail
of ``multimessenger'' astronomy) would solidify the compact binary
coalescence model of sGRBs. However, the bulk of sGRBs have been found
at redshifts $z>0.1$~\cite{Berger2014}, i.e., at luminosity distances
$D_L \gtrsim 460$ Mpc (assuming standard $\Lambda$CDM cosmology) and
hence outside the aLIGO NSNS horizon. Also the most recent estimates
of sGRB rates suggest a rate of $8/\rm
yr$~\cite{Fong2015ApJ...815..102F} within the aLIGO NSNS horizon of
$\sim 200$ Mpc\footnote{The aLIGO BHNS horizon with masses $10M_\odot$
  for the BH and $1.4M_\odot$ for the NS is $\sim 900$Mpc, but unless
  the BH is rapidly spinning or the NS is sufficiently puffy these
  systems may not be able to power sGRBs (see Sec.~\ref{BHNShydro}
  below).}. Thus, a solid identification of an sGRB with a GW signal
may require either several years of LIGO/Virgo observations or a lucky
nearby sGRB that happens to point toward the Earth.  However, as rate
estimates typically have substantial uncertainties, we might even have
to wait for third-generation GW observatories until such an
identification may be possible. But, until this happens, a
theoretical/computational study of compact binary mergers with the aid
of numerical relativity is an important avenue to gaining a better
understanding of these systems as jet engines for sGRBs and for
inferring intrinsic properties/parameters of the engines from EM
observations alone.

Recent years have witnessed a growing number of compact binary
simulations in full general relativity with different levels of
sophistication and realism. Although these simulations have
contributed to improving our understanding of BHNS and NSNS mergers,
we are still far away from constructing a complete
theoretical/computer generated sGRB model starting from the inspiral
and merger all the way to jet acceleration and emergence of the
gamma-ray burst. The existence of such a model would solidify compact
binary mergers as viable sGRB engines on theoretical grounds, and
would, in principle, allow for the extraction of the progenitor binary
parameters from the gamma-ray signal even in the absence of a GW
counterpart. Moreover, a complete theoretical model of sGRBs would
dictate the time lag between the peak GW amplitude and the gamma-ray
burst, and thereby would better inform triggered GW searches. However,
it is hard to envision that such an sGRB model is achievable in the
foreseeable future because of the very high-resolution requirements to
capture the relevant magnetic effects, the disparity of length and
time scales involved in the problem and because of the very difficult
neutrino transport problem involved following merger. For now, it
seems that the combination and coupling of different codes simulating
different phases of the evolution of an sGRB engine offer the only
plausible route for building such a complete model of an sGRB. But,
even this approach is several years away from being realized.

There exists a vast literature on theoretical/computational methods
for modeling the different phases of an sGRB engine, however, here, we
will focus on the very first stage, i.e., the formation of the BH-disk
engine from the inspiral and merger of compact binaries involving
neutron stars and the early launch of jets. This phase of the sGRB
engine requires the field of numerical relativity, and hence this
review is centered on the status of state-of-the-art
(magnetohydrodynamic) simulations of compact binary mergers as sGRB
engines. Since there exists only little work on simulations of compact
binaries with a neutron star component in modified gravity
theories~\cite{2013PhRvD..87h1506B,2014PhRvD..89d4024P,ShibtaST2014PhRvD..89h4005S,2015PhRvD..91h4038P},
the focus of this review will be on simulations in full GR, and in
particular we will highlight the latest developments in this subfield.

The remainder of the paper is structured as follows. In
Section~\ref{challenge} we review the computational challenge involved
in modeling compact binaries and discuss the equations that govern
their dynamical evolution; in Sec.~\ref{BHNS} we review recent results
obtained from state-of-the-art simulations of binary BHNSs and in
Sec.~\ref{NSNS} recent results from state-of-the-art simulations of
binary NSNSs. We conclude in Sec.~\ref{other} with a brief discussion
and list of open questions. Unless otherwise specified, below we adopt
geometrized units, where $G=c=1$.

\section{The challenge}
\label{challenge}

The NSNS and BHNS inspiral and merger problem is a multi-scale and
multi-physics one. The range of length and time scales in the problem
in conjunction with the large number of non-linear partial
differential equations one must solve, makes modeling these binaries a
very challenging task.

\subsection{Length and time scales}

The range of length and times scales involved in this problem spans
over 3-4 orders of magnitude. For example, to reliably model the
inspiral over the last few orbits and to resolve the neutron star(s)
and/or black hole requires at least 100 grid points across the radius
of each object. This implies that for a typical NS with radius $\sim
10$km, the grid spacing must be $\Delta x \sim 0.1$km. However,
computing reliable gravitational waveforms, requires that the GW
extraction be done at a radius of $\sim 100M \sim 450(M/3M_\odot)$km
or greater. Here $M$ is the binary total mass. To properly capture
hydromagnetic effects during an NSNS merger and in the post-merger
BH-disk system that forms, a grid-spacing of order $10$m seems to be
necessary~\cite{Zrake2013ApJ...769L..29Z,2015PhRvD..92l4034K,kskstw15}. In
other words, the length scales span 3-4 orders of magnitude from
inspiral through merger. Therefore, some level of mesh refinement is
necessary, and all modern numerical relativity codes adopt adaptive
mesh refinement. If a jet emerges shortly after merger, tracking its
evolution until it achieves terminal Lorentz factor, requires that one
be able to follow the jet to distances of several hundreds of
thousands to millions of $M$ from the engine, where the jet may
typically reach its terminal Lorentz factor (see
e.g.~\cite{Tchekhovskoy01082008}). The sparsity of length scales
demonstrates quite clearly how difficult it becomes to model an sGRB
from first principles, but see~\cite{Duffel2015ApJ...806..205D} for a
method that can be used if the jet decouples from the central engine,
which occurs when the jet becomes supersonic.

Some fundamental time scales involved in the BHNS problem are the
timestep, NS dynamical time scale, the inspiral time scale from the
initial orbital separation, and the incipient jet emergence time scale.

Due to the Courant limitation, the timestep must be of order $\Delta t
\simeq 0.5 \Delta x \simeq 1.5\times 10^{-4} (\Delta x/100{\rm m})$
ms. The NS dynamical time scale is
\begin{equation}
t_d = 2\pi\sqrt{\frac{R_{\rm NS}^3}{M_{\rm NS}}}\simeq 0.5 \bigg(\frac{R_{\rm NS}}{10\rm km}\bigg)^{3/2}\bigg(\frac{M_{\rm NS}}{1.35M_\odot}\bigg)^{-1/2}\rm ms,
\end{equation}
where $R_{\rm NS}$ is the NS radius and $M_{\rm NS}$ the NS mass.

The inspiral time to merger from an initial orbital separation is
dictated by the gravitational wave time scale, which in the quadrupole
approximation is given by
\begin{equation}
t_{\rm merge} = \frac{5}{16}\frac{a^4}{M^3\zeta}\simeq
3000\bigg(\frac{a}{10M}\bigg)^4 \zeta^{-1}M\simeq 45
\bigg(\frac{a}{10M}\bigg)^4 \bigg(\frac{M}{3M_\odot}\bigg)\zeta^{-1} \rm ms
\end{equation}
where $a$ is the orbital separation, $\zeta=4\eta=4q/(1+q)^2$, with
$\eta$ the symmetric mass ratio, and $q=M_1/M_2$ the usual binary mass
ratio. Note, that for an equal-mass binary $\zeta=1$.

The jet emergence time is of order $100(M/3M_\odot)$ ms after
merger~\cite{prs15} and could be longer (see below). Thus, evolving
from inspiral through jet launching, requires a total evolution of
$\sim 100-150$ ms, which implies that of order $10^6$ time steps are
not atypical in these simulations.

In an NSNS merger scenario, a BH may not form immediately after
merger, and instead, a differentially rotating hypermassive neutron
star (HMNS) may be the merger outcome. A HMNS is a transient
object. It will undergo ``delayed collapse'' on a secular time scale
(see e.g.~\cite{BaShSh,dlsss06a}), and then form the BH-disk engine
that may power a sGRB. An HMNS is supported against collapse by a
combination of the additional centrifugal support due to differential
rotation and the extra thermal pressure generated by shock heating
(see e.g.~\cite{PhysRevLett.107.051102,Paschalidis:2012ff}). The
BH-disk system will then likely form on an Alfv\'en time $t_{\rm Alf}$
(the time scale for the braking of the differential of the HMNS by
magnetic fields~\cite{Shapiro:2000zh}) or the GW time scale (the time
over which angular momentum is carried away through GW emission from a
non-axisymmetric HMNS) $t_{\rm GW}$, and possibly even the cooling
time $t_\nu$ due to neutrino
emission~\cite{PhysRevLett.107.051102}. These time scales are given
by~\cite{Paschalidis:2012ff}
\labeq{talfven}{ t_{\rm Alf} \approx 30 \bigg(\frac{R}{20 \rm
    km}\bigg)^{-1/2} \bigg(\frac{M}{2.8
    M_{\odot}}\bigg)^{1/2}\bigg(\frac{B}{10^{15.5} \rm G}\bigg)^{-1} \rm
  ms,
}
where $R$ is the characteristic radius of the HMNS, $M$ the mass of
the HMNS, which approximately equals the total mass of the NSNS, and
$B$ a typical value of the magnetic field strength of the HMNS. The GW
time scale is
\labeq{tGW}{ t_{\rm GW} \approx 200 \bigg(\frac{e}{0.75}\bigg)^{-2}
  \bigg(\frac{R}{20 \rm km}\bigg)^{4} \bigg(\frac{M}{2.8
    M_{\odot}}\bigg)^{-3} \rm ms, }
where $e$ is the HMNS ellipticity and where we adopted a value of a
plausible bar-like configuration. The estimated $t_{\rm GW}$ is
comparable to the GW time scale inferred by numerical relativity NSNS
simulations (see e.g.~\cite{Rezzolla:2010fd}). Finally, the cooling
time scale is estimated as follows (see
also~\cite{RuffertJanka1996A&A...311..532R} for an estimate accounting
for trapping effects in deformed objects)
\labeq{tCool2}{ t_{\rm \nu} \approx 1 \bigg(\frac{M}{2.8
    M_{\odot}}\bigg)\bigg(\frac{R}{20 \rm
    km}\bigg)^{-1}\bigg(\frac{E_\nu}{15\rm MeV}\bigg)^2 \rm s, }
where $E_\nu$ is the rms energy of the emitted neutrinos. Thus, if the
HMNS is primarily supported by thermal pressure, as has been argued to
be the case in a simulation in~\cite{PhysRevLett.107.051102}, one may
need to evolve for 1s to form the BH disk engine in NSNS merger
scenario, which implies of order $10^7$ timesteps. As a result,
numerical relativity studies of BH formation in NSNS mergers have
focused only on cases where a BH forms within $\sim 100$ ms following
merger (see Appendix of~\cite{Rezzolla:2010fd} for the longest NSNS
hydrodynamic simulation in full GR).

\subsection{Equations}

The multi-scale nature of the inspiral and merger problem of compact
binaries is one aspect of the challenge numerical relativity
simulations face. The multi-physics nature of the problem is another
one. Moreover, compact objects are inherently relativistic, requiring
that these simulations be performed in full general relativity, and
thereby complicating the task of the modeler even further. The term
``multi-physics'' implies that a large number of equations must be
solved. In particular, the equations describing compact binaries
involving neutron stars are (see e.g.~\cite{BSBook}) the following:

a) The Einstein equations which govern the evolution of the spacetime, and
are given by
\begin{equation}
G_{\mu\nu} = 8\pi T_{\mu\nu},
\end{equation}
where $G_{\mu\nu}$ is the Einstein tensor and $T_{\mu\nu}$ the matter
stress-energy tensor. The Einstein equations are 10 second-order,
non-linear partial differential equations (PDEs). 

b) The energy-momentum and radiation transport equations which govern
the evolution of the matter and radiation and are given by
\begin{eqnarray}
\nabla_\mu T^{\mu\nu} & = & -\nabla_\mu R^{\mu\nu}, \\
\nabla_\mu R^{\mu\nu} & = & -G^\nu,
\end{eqnarray}
where $R_{\mu\nu}$ is the radiation stress tensor and $G^\nu$ the
radiation four-force density. These form a set of another 8
PDEs. 

c) Maxwell's equations which govern the evolution of the
electromagnetic fields, and are given by
\begin{eqnarray}
\nabla_\mu F^{\mu\nu} &=& -4\pi J^{\nu}, \\
\nabla_\mu{} ^{*}F^{\mu\nu} & = & 0,
\end{eqnarray}
where $F_{\mu\nu}$ is the electromagnetic tensor and $^{*}F^{\mu\nu}$
its dual. Maxwell's equations add another 8 PDEs. 

d) The baryon number conservation (or continuity) equation
\begin{equation}
\nabla_\mu (\rho_0u^\mu) = 0.
\end{equation}
These add up to a total of 27 coupled PDEs in 3 spatial plus 1
temporal dimensions.  In fact, the radiation transport equation has an
additional 3 dimensions (two angular and the radiation frequency)
making it a 6+1 dimensional problem. This system of equations must
also be supplemented with a microphysical, hot, nuclear EOS and, in
the general case, with an Ohm's law for the current, which both add even
more to the complexity of the problem. Note that the above counting
does not even account for the lepton number conservation equations nor
does it account for the fact that these equations are not in a form
amenable for numerical integration, and that standard formulations of
these equations used in numerical simulations involve many more
coupled PDEs (see
e.g.~\cite{Alcubierre2008book,BPB2008book,BSBook,Gou2012book,Shi2015book}
for textbooks).

The large number of non-linear coupled equations is not the only
challenge a numerical relativist faces. Other unique challenges
involve curvature singularities (infinities) that one must treat
properly when modeling BH spacetimes, and that in general relativity
only gauge independent quantities are meaningful, which makes
extracting physical information from these simulations a non-trivial
task. However, many of these difficulties have been overcome over the
years by creative theoretical and computational methods. For a
comprehensive description of such methods
see~\cite{Alcubierre2008book,BPB2008book,BSBook,Gou2012book,Shi2015book}.
Multiple codes have been developed that solve the full Einstein
equations coupled to all or a subset of the remaining equations
adopting various degrees of realism and
approximations~\cite{DLSS,Giacomazzo:2007ti,Etienne:2010ui,Etienne:2011re,Farris:2012ux,Sekiguchi2010CQGra..27k4107S,Thierfelder2011PhRvD..84d4012T,Kiuchi2012PhRvD..86f4008K,code_paper,Palenzuela:2012my,ET2012CQGra..29k5001L,Dionysopoulou2013,Galeazzi2013PhRvD..88f4009G,Deaton:2013sla,Radice2014CQGra..31g5012R,Neilsen2014PhRvD..89j4029N,Muhlberger2014,Foucart2015PhRvD..91l4021F,IllinoisGRMHD}. These
codes have been applied to the study of both BHNS and NSNS mergers.

\section{BHNS mergers}
\label{BHNS}

It has been about 10 years since hydrodynamic simulations of BHNS
mergers in full general relativity have become
routine~\cite{BSS,TBFS05,FBSTR,FBST,SU14774,eflstb08}. BHNS
simulations combine all the physical challenges encountered in
magnetohydrodynamics, such as magnetized shock discontinuities, with
those of vacuum black holes, namely a curvature singularity. The
latter is the primary reason why the advancement of BHNS simulations
lagged compared to NSNS simulations, and were made possible only after
the breakthrough BHBH vacuum simulations
of~\cite{FP1,Baker:2005vv,RIT1}. For a comprehensive review surveying
BHNS simulations see~\cite{st11}. Here we only review studies of these
systems in full GR and only relevant to sGRBs.

\subsection{Hydrodynamic Simulations}
\label{BHNShydro}

Primordial BHNS binaries are likely quasicircular because GW emission
tends to circularize the orbit~\cite{pm63}. These binaries are also
anticipated to involve irrotational neutron stars, because the tidal
synchronization time scale exceeds the inspiral time
scale~\cite{Bildsten1992ApJ...400..175B}. 

Motivated by the above conclusions, the first studies of BHNS mergers
testing the viability of BHNSs as sGRB engines, focused on
hydrodynamic simulations of quasicircular, irrotational binaries, with
the goal of determining the parameter space within which an
appreciable accretion disk may form outside the BH following the NS
tidal disruption. Having a BH-disk remnant is important to power an
sGRB because even a small fraction of the accretion power can account
for typical sGRB luminosities. The characteristic sGRB durations and
luminosities dictate the amount of matter that an accretion disk
should have as follows: Assuming an efficiency $\epsilon$ for
converting the accretion luminosity $\dot M c^2$ to gamma-ray
luminosity $L_\gamma$, the matter accretion rate onto the BH engine
becomes $\dot M =\epsilon^{-1} L_\gamma/c^2$.  The disk lifetime,
which provides the sGRB fuel, should be of order the typical sGRB
duration $t_{\rm sGRB}$~\cite{LeeRamirezRuiz2007}. Thus, the disk mass
can be estimated as $M_{\rm disk}\sim \dot M \times t_{\rm sGRB}$,
yielding
\begin{equation}
\label{Mdiskreq}
M_{\rm disk} \sim 0.1M_\odot\bigg(\frac{\epsilon}{0.01}\bigg)^{-1} \bigg(\frac{L_\gamma}{10^{51} \rm erg/s}\bigg)\bigg(\frac{t_{\rm sGRB}}{2\rm s}\bigg)
\end{equation}
Hence, $\sim 8\%$ of the NS rest mass (assuming $M_{\rm
  NS}=1.3M_\odot$ and 1\% efficiency) should remain outside the BH
following merger, in order to power a 2s-long, $10^{51}$ erg/s sGRB
(the upper end on duration of sGRBs). However, the average sGRB
duration is 0.3s~\cite{Kouveliotoi1993ApJ...413L.101K}, thus a $\sim
0.015M_\odot$ disk suffices for most cases. 

But, forming a disk outside the BH is not trivial, because for this to
occur the NS must be tidally disrupted outside the BH's innermost
stable circular orbit (ISCO). To understand why this is difficult one
can derive an order of magnitude estimate for the tidal disruption
radius $a_d$ of the NS by equating the BH tidal force to the NS
gravitational force on the NS surface~\cite{illinoisBHNS}, which
yields
% C = M_NS/R_NS => R_NS= M_NS/C = M_BH/qC
% because q = M_BH/M_NS => M_NS = q/M_BH
\begin{equation}
\label{ad}
\frac{2M_{\rm BH}R_{\rm NS}}{a_d^3} \simeq \frac{M_{\rm NS}}{R_{\rm NS}^2} \Longrightarrow
\frac{a_d}{M_{\rm BH}} \simeq 3 \bigg(\frac{C}{0.2}\bigg)^{-1}\bigg(\frac{q}{7}\bigg)^{-2/3},
\end{equation}
where $C=M_{\rm NS}/R_{\rm NS}$ is the NS compaction, and $q=M_{\rm
  BH}/M_{\rm NS}$ the BH to NS mass ratio and a value of 7 was adopted
since it is anticipated to be the most probable value for primordial
BHNS binaries~\cite{bdbofh10}. Recall that the ISCO radius is $6M_{\rm
  BH}$ for a non-spinning BH, $M_{\rm BH}$ for a maximally spinning BH
and $3M_{\rm BH}$ for a BH with dimensionless spin parameter
$\chi=a_{\rm BH}/M_{\rm BH} \sim 0.78$~\cite{RISCO1972}. The simple Newtonian
estimate of Eq.~\eqref{ad} demonstrates how difficult it is to disrupt
a NS outside the ISCO of a slowly spinning BH. Therefore, an
appreciable disk may form following the NS tidal disruption only for
relatively small mass ratios and either if the BH is highly spinning
and/or the NS is not very compact (smaller $C$).

\begin{figure}
  \centering \includegraphics[width=0.5\textwidth]{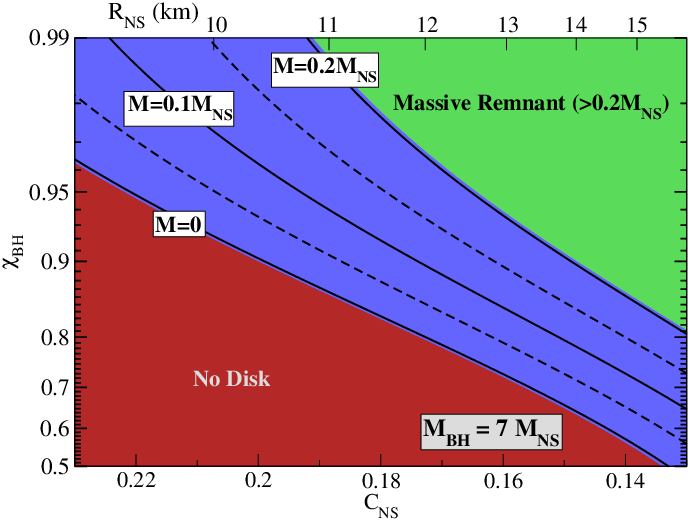}
  \caption{Contours of disk mass left outside the BH following a BHNS
    merger at fixed mass ratio q=7. Here $C_{\rm NS}$ is the NS
    compaction and $\chi_{\rm BH}$ the BH dimensionless spin. The NS
    radius shown on the top horizontal axis assumes a NS gravitational
    mass of $1.4M_\odot$. Figure 7 from~\cite{2012PhRvD..86l4007F}.
  \label{fig:bhnsdisk}}
\end{figure}

Multiple numerical relativity investigations have studied how much
matter remains outside the BH following a quasicircular, irrotational
BHNS merger, and the results were compiled
in~\cite{2012PhRvD..86l4007F}, where the following EOS-independent
fitting formula was proposed for predicting the mass $M_{\rm disk}$
left outside the BH to form a disk that may power a sGRB
\labeq{BHNSdiskmass}{
\frac{M_{\rm disk}}{M_{\rm NS}}=0.415q^{1/3}(1-2C)-0.148R_{\rm ISCO}/R_{\rm NS},
}
where $R_{\rm ISCO}$ denotes the ISCO radius of the initial
BH. However, note that the applicability of this formula is restricted
to dimensionless BH spins $\chi \lesssim
0.9$~\cite{Lovelace2013CQGra..30m5004L}. Using
Eq.~\eqref{BHNSdiskmass} one can plot contours of disk mass as a
function of compaction, and BH spin for a given mass ratio. This is
shown in Fig.~\ref{fig:bhnsdisk} for the most probable value of the
mass ratio for primordial BHNS binaries~\cite{bdbofh10}. The main
conclusion drawn from Fig.~\ref{fig:bhnsdisk} is that \emph{at
  ``realistic'' mass ratios and values of the NS compaction of
  $C=0.18$ suggested by
  observations~\cite{Steiner2010ApJ...722...33S}, the disk mass will
  be $\gtrsim 0.1M_\odot$ only if the BH dimensionless spin is
  $\chi\gtrsim 0.8$, which could be a very tight
  constraint.}

The previous conclusion holds for quasicircular BHNSs in which the NS
is irrotational.  While quasicircular binaries probably dominate the
BHNS merger rates in the Universe, some recent
results~\cite{Kocsis:2011jy,lee2010,Samsing:2013kua} indicate that in
dense stellar regions, for example galactic nuclei and globular
clusters (GCs), compact binaries can form through single-single
(dynamical capture) and binary-single (exchange) interactions, and
merge with substantial eccentricities. Rates of these eccentric
mergers are highly uncertain, but the optimistic ones can be up
$\sim100\rm{yr}^{-1}\ Gpc^{-3}$~\cite{ebhns_letter,bhns_astro_paper},
i.e., comparable to the lower bound on the estimated merger rate for
primordial binaries~\cite{KBKOW,deMink2015ApJ...814...58D}. Another
important aspect of GC neutron stars is that more than 80\% of pulsars
residing in GCs have periods less than 10 milliseconds (i.e. are
millisecond pulsars), and hence compact binary mergers with at least
one neutron star component occurring in GCs could involve rapidly
spinning neutron stars~\cite{EPP2015,PEPS2015,EPPS2016}. High NS spin
makes the star less bound, increasing the tidal disruption radius, and
both prograde rapid NS spin and orbital eccentricity move the
effective innermost stable orbit (ISO) inward allowing for the NS to
be tidally disrupted outside the ISO even when a non-spinning BH is
involved.

Motivated by the above, hydrodynamic simulations in full GR of
dynamical capture BHNS mergers have been performed
in~\cite{ebhns_letter,bhns_astro_paper} with non-spinning NSs and
in~\cite{EPP2015} with spinning NSs for $q=4$. Depending on the value
of the periapse distance during the final encounter, the amount of
disk mass for non-spinning BHs found ranges from $\sim 1\%- 10\%$ of
the NS rest mass for moderate stiffness EOSs ($C=0.17$), and it can be
up to $15\%$ for stiff EOSs ($C=0.13$). By contrast,
Eq.~\eqref{BHNSdiskmass} for a $q=4$ quasi-circular BHNS predicts no
mass outside the BH for $C \gtrsim 0.135$ and only $2.6\%$ for
$C=0.13$. Therefore, dynamical capture BHNS mergers as may arise in
GCs are viable progenitors for sGRBs and can potentially generate
BH-disk engines more easily than quasi-circular mergers with the same
BH:NS mass ratio and initial black hole spin.

Showing that a compact binary system can form a BH-disk engine is the
first step in demonstrating theoretically the viability of compact
binaries as progenitors of sGRBs (in the hyperaccreting and
jet-launching BH model). The second crucial step is to show that these
BH-disk engines can launch jets that can be accelerated to a Lorentz
factor $\Gamma_L \gtrsim 100$, which is a crucial ingredient in the
fireball model of sGRBs~\cite{Meszaros:2006rc}. The most popular
mechanisms invoked for launching and accelerating jets are either
magnetic fields~\cite{BZeffect} or neutrino
annihilation~\cite{Mochkovitch1993Natur.361..236M,Alloy2005A&A...436..273A},
but it also could be that a combination of the two is necessary. Most
simulations in full GR to date have focused on the magnetic launching
mechanism mainly because the neutrino annihilation mechanism requires
proper treatment of the neutrino transport equation which is far from
an easy task. Therefore, next we will summarize the status of
magnetohydrodynamic simulations of BHNS mergers in full GR.

\begin{figure}
  \centering
  \includegraphics[width=0.7\textwidth]{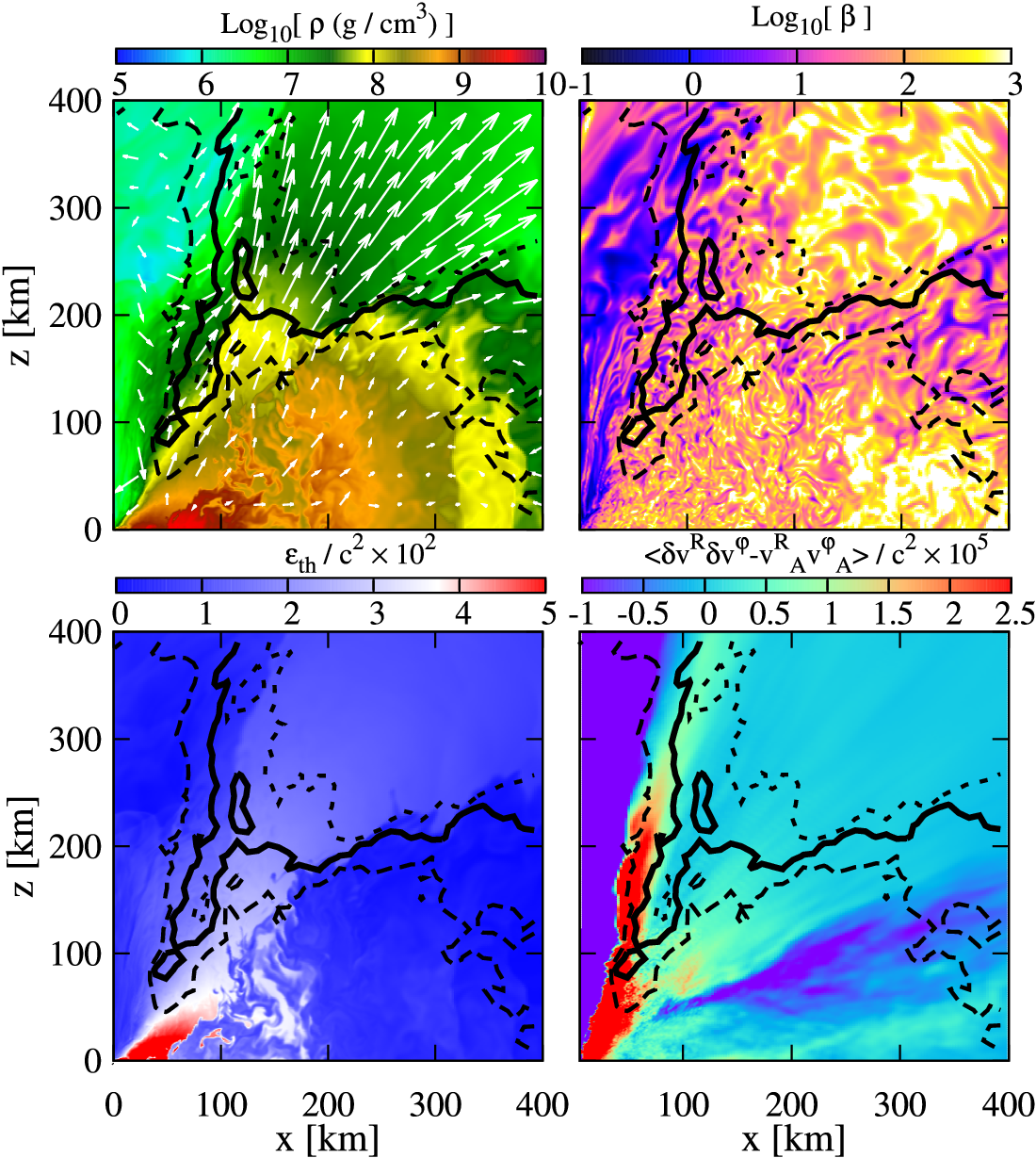}
  \caption{Top left: rest-mass density with velocity arrows. A wind
    outflow is observed, but no collimated outflow is found within
    funnel walls formed by the wind. Top right: plasma beta
    parameter. Bottom left: thermal specific internal energy. Bottom
    right: sum of Maxwell and Reynolds stress (bottom-right). All
    profiles are shown on the $x-z$ plane at $t\sim 50$ ms. Figure
    5 from~\cite{kskstw15}.
  \label{fig:ksks}}
\end{figure}

\subsection{Magnetohydrodynamic Simulations}

The combination of a spacetime singularity, hydrodynamic shocks and
the presence of magnetic fields renders simulations of BHNS systems
extremely challenging. As a result only few magnetohydrodynamic (MHD)
simulations of BHNS systems in full GR have been performed to date.

\begin{figure}
  \centering
  \includegraphics[width=0.4\textwidth]{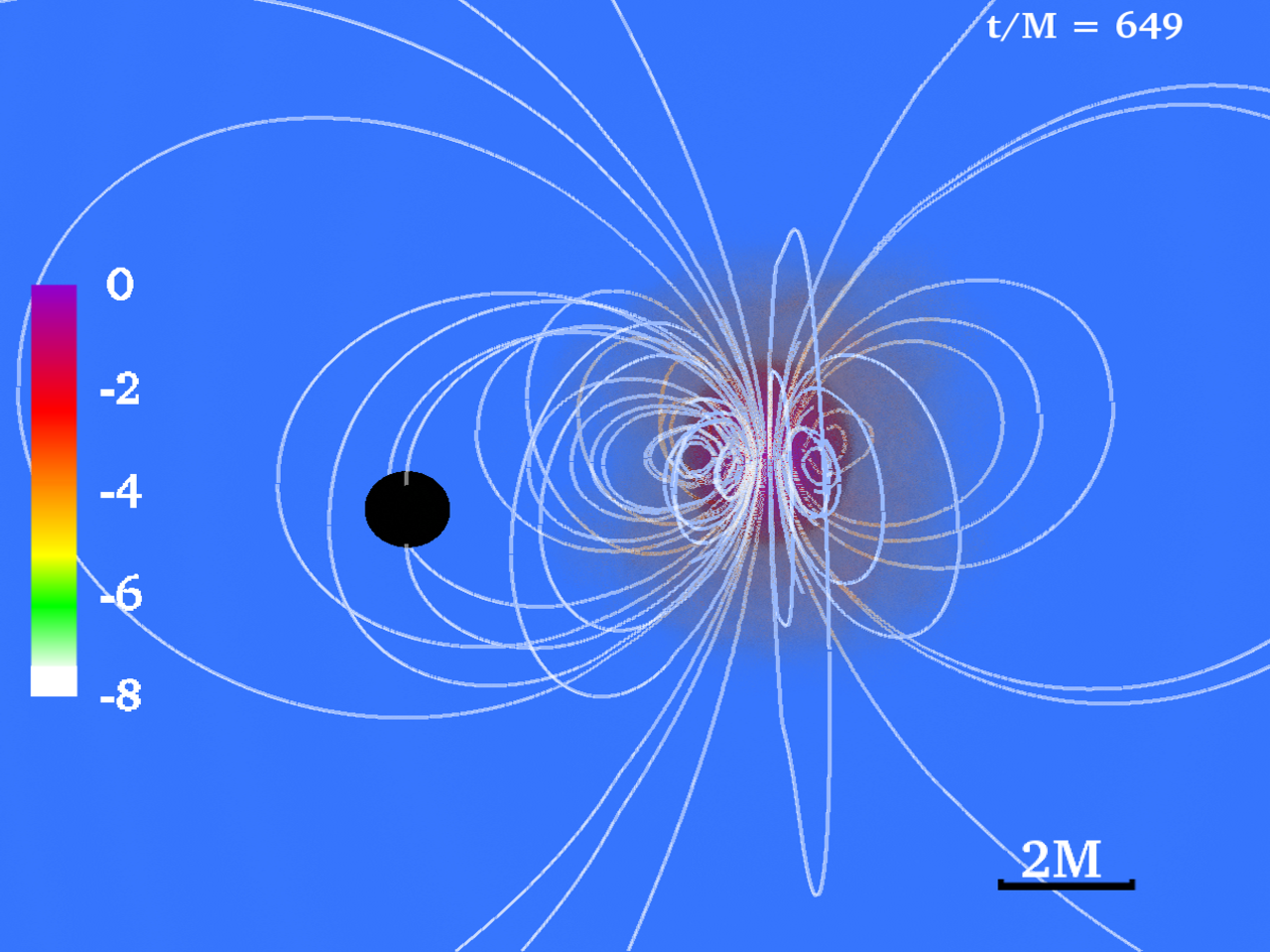}
  \includegraphics[width=0.4\textwidth]{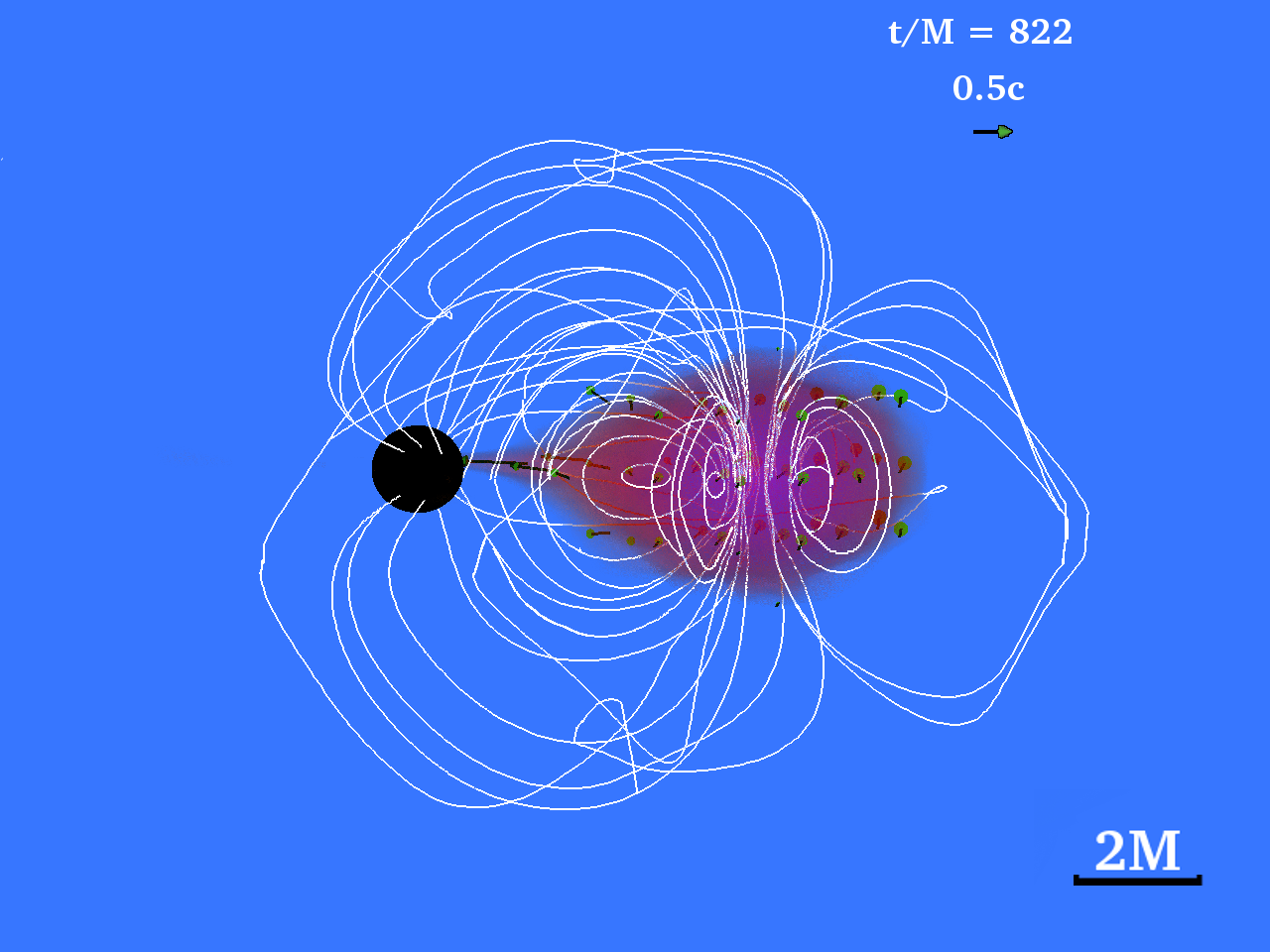}
  \includegraphics[width=0.4\textwidth]{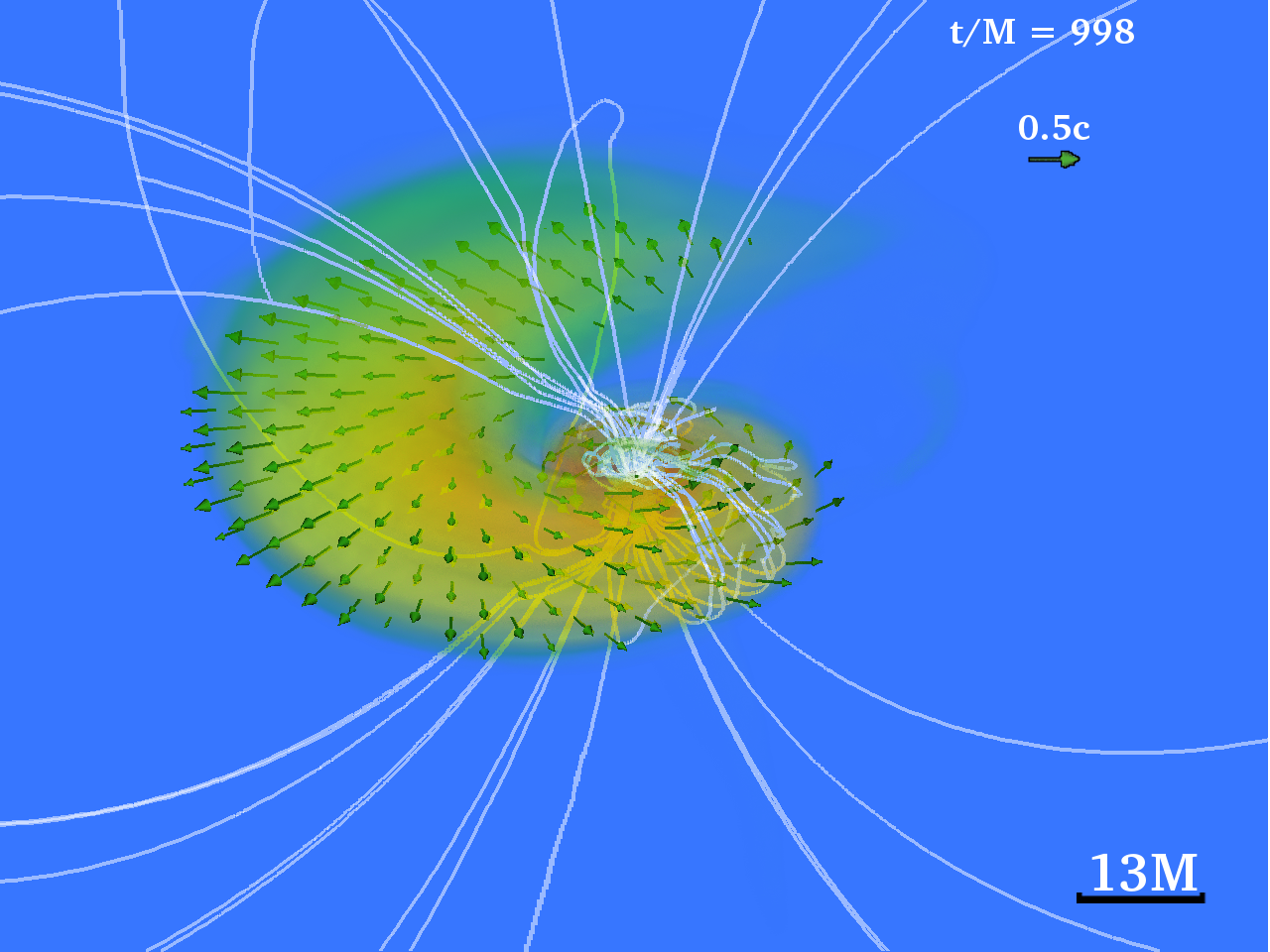}
  \includegraphics[width=0.4\textwidth]{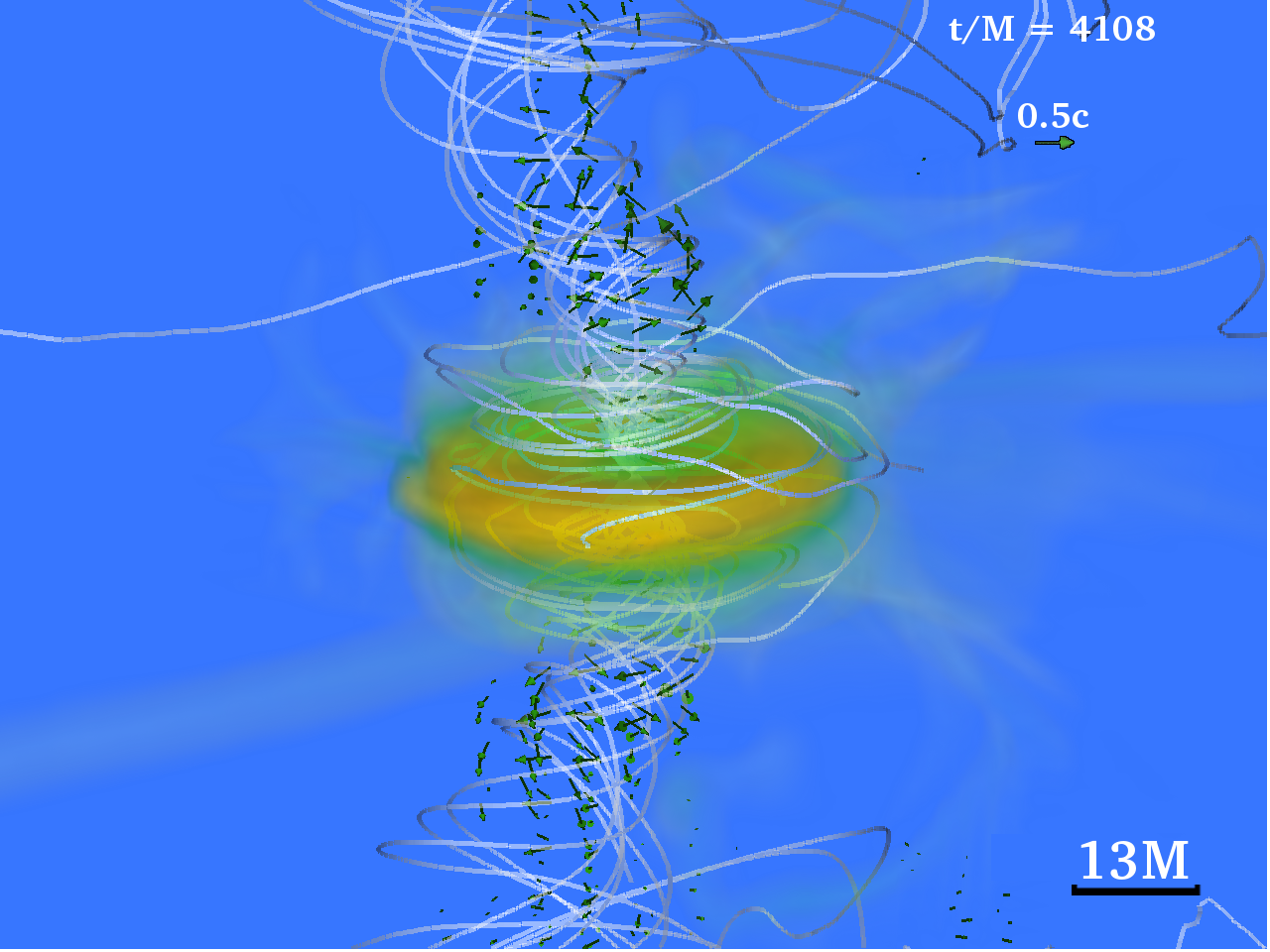}
  \includegraphics[width=0.4\textwidth]{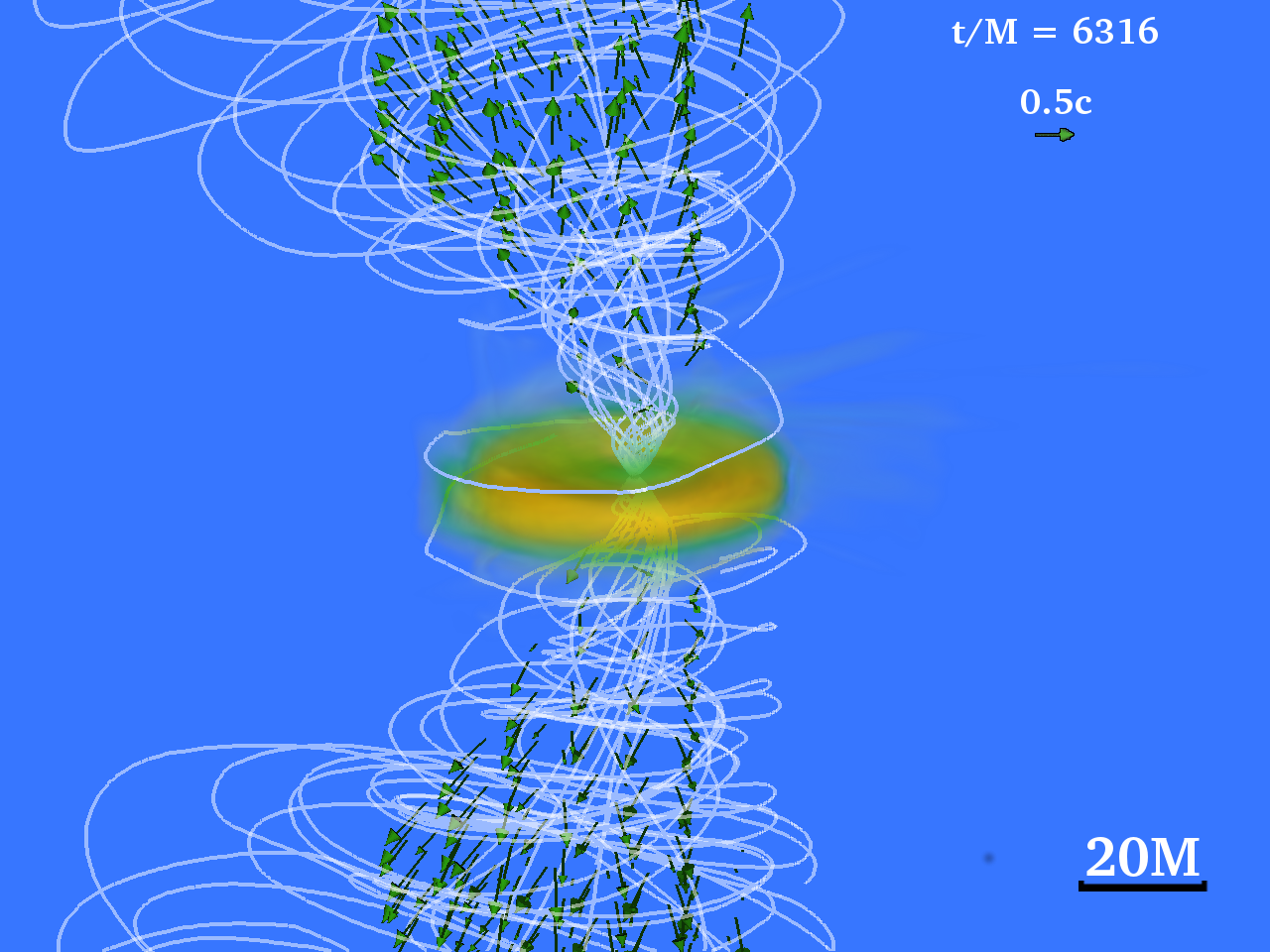}
  \includegraphics[width=0.4\textwidth]{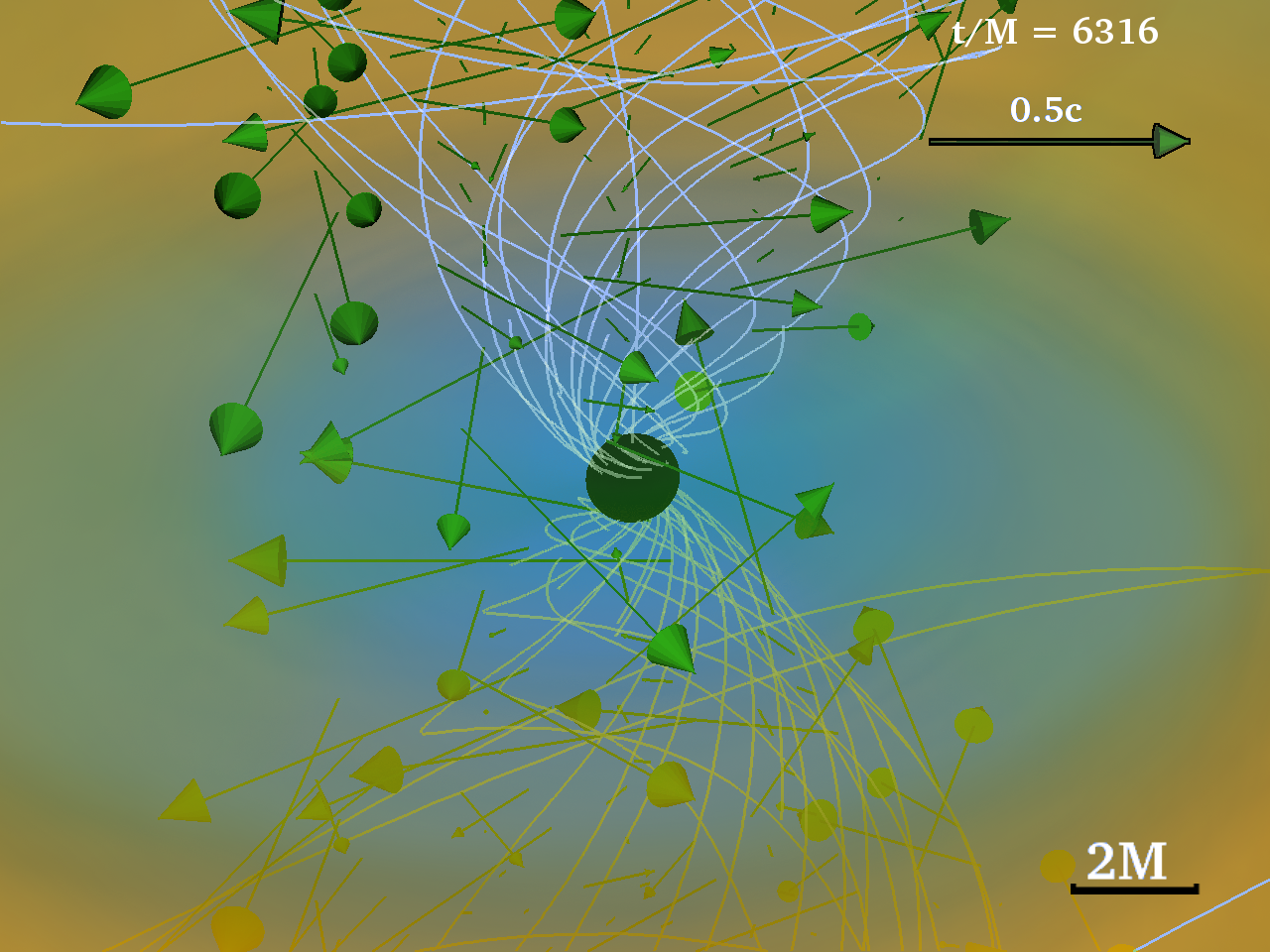}
  \caption{Volume rendering of the rest-mass normalized to its initial
    maximum value~$\rho_{0,\rm max}=8.92\times
    10^{14}(1.4M_\odot/M_{\rm NS})^2\rm g\ cm^{-3}$ (log scale) at
    select times. Arrows indicate matter velocities and white lines
    the magnetic field lines. The bottom panels show the system after
    an incipient jet has been launched. Here, ${\rm M}=2.5\times
    10^{-2}(M_{\rm NS}/1.4M_\odot)$ms$=7.58(M_{\rm
      NS}/1.4M_\odot)$km. Figure 1 from~\cite{prs15}.
  \label{fig:Binit}}
\end{figure}

The MHD simulations in full GR performed in~\cite{cabllmn10} reported
the formation of a viable BH-disk engine, but no jets were found.  The
studies of magnetized BHNS mergers in full GR as sGRB engines carried
out in~\cite{Etienne:2011ea,Etienne:2012te} probed an expanded part of
the BHNS parameter space. The most promising case in terms of the
amount of disk mass left outside the BH was a 3:1 mass ratio with an
initially spinning BH $\chi = 0.75$. This configuration results in
$\sim 10\%$ of the initial neutron star rest mass forming an accretion
disk around the remnant BH with dimensionless spin of $\sim 0.85$. In
these studies the initial neutron star was seeded with a dipole
magnetic field confined entirely in the interior of the star and
equatorial symmetry was imposed during the evolution. The results
about jet emergence were null, even when the strength of the initial
magnetic field seeded in the NS was much stronger when compared to
standard inferred values of pulsar magnetic fields. The initial
explanation for the lack of a jet was that the resolution was too low
to resolve the wavelength of the fastest growing mode of the
magnetorotational instability (MRI) in the resulting disk. However,
even when lifting the equatorial symmetry and increasing the
resolution to resolve the MRI wavelength by 10 grid points -- the rule
of thumb for capturing the basic magnetic field growth due to
MRI~\cite{MRI_RESOLVED_TO_10_GRIDPOINTS_PAPER} -- again no jet
emerged.

Similar results were reported in~\cite{kskstw15} who performed MHD
simulations of BHNSs in full GR adopting higher resolution. The
initial configuration consisted of a quasiequilibrium binary, with an
irrotational NS of mass $1.35M_\odot$, mass ratio $q=4$, and BH spin
$\chi=0.75$. At $t=10$ ms after merger a BH-disk system forms with
$\sim 0.13M_\odot$ in the disk. But, after $\sim 50$ ms of evolution
the authors of~\cite{kskstw15} found no jets. However, they reported a
wind outflow (see Fig.~\ref{fig:ksks}), and the emergence of a large
scale, poloidal component of the magnetic field.

The lack of jets from these BHNS simulations was very puzzling and
came as a surprise, because for over 15 years GRMHD accretion studies
onto BHs in fixed spacetime demonstrated that jets naturally arise in
these scenarios (see e.g.~\cite{lrr-2013-1} for a review). However, as
it turns out this is not always the case.

For a magnetized accretion disk with magnetic fields initially
confined in the disk interior, a jet can be launched and sustained
only if the initial magnetic field is such that a net poloidal
magnetic flux is accreted onto the
BH~\cite{GRMHD_Jets_Req_Strong_Pol_fields}. For example, starting with
purely toroidal magnetic fields in the disk no jets are launched. This
result naturally explains why no jets were found
in~\cite{Etienne:2011ea,Etienne:2012te}.  In particular, following
tidal disruption of the NS by the BH, the bulk of the magnetic field
flux flows instantaneously into the BH, and the magnetic field
remaining outside is wound into an almost purely toroidal
configuration. In addition, the residual poloidal component of the
magnetic field in the disk does not have a consistent vertical sign in
the sense of~\cite{GRMHD_Jets_Req_Strong_Pol_fields}, hence no jets
can be launched.

The missing ingredient for jet launching was finally identified
in~\cite{prs15}, who made the realization that all early GRMHD BHNS
studies used magnetic fields confined in the interior of the
NS. Pulsars suggest that a more realistic magnetic field configuration
is a dipole extending from the NS interior out to the
exterior~\cite{prs15}. While it appears this would be a trivial
change, it is not, because it is very challenging to adopt an ideal
MHD code to evolve regions where the magnetic field energy density
dominates over the rest-mass energy density, as in a pulsar
magnetosphere. To overcome this obstacle, the authors designed novel
initial conditions, which would capture only one aspect of a
magnetosphere, namely magnetic-pressure dominance, but not
magnetic-energy density dominance. These new initial conditions
allowed the evolution of exterior magnetic fields with an ideal GRMHD
code and using a sequence of simulations from weakly to highly
magnetic-pressure dominated exteriors, the authors were able to test
whether their results were invariable in this sequence.

Following the above approach the authors followed the BHNS encounter
through tidal disruption and BH-disk formation. As in the past the
magnetic field in the disk interior was predominantly
toroidal. However, the no-go conditions for jet launching
of~\cite{GRMHD_Jets_Req_Strong_Pol_fields} could now be evaded because
of the existence of an exterior magnetic field. In particular, a
poloidal magnetic field component was present throughout the evolution
(see Fig.~\ref{fig:Binit}). The authors reported that soon after the
violent accretion episode and disk settling, the magnetic field above
the BH poles was amplified from $\sim 10^{13}$ G to $\sim 10^{15}$ G
primarily via winding [see Eqs.~\eqref{Bdisk},~\eqref{Bthres} below
  for an argument of why such magnetic-field strengths arise naturally
  when jets are launched following compact binary mergers]. This
strong magnetic field finally drove an incipient jet about $\sim 100$
ms following the BHNS merger (see Fig.~\ref{fig:Binit}). The result
was the same for all magnetospheric conditions ranging from moderate
to high magnetic-pressure dominance (magnetospheric plasma parameter
$\beta_{\rm ext}=P_{\rm gas}/P_{\rm mag}\in [0.01,0.1]$), with larger
initial $\beta_{\rm ext}$ resulting in delayed jet launching. The disk
lifetime found was 0.5 s, and the jet Poynting luminosity $10^{51}$
erg/s, which are both entirely consistent with typical sGRBs. The
Poynting luminosity was also consistent with the Blandford-Znajek (BZ)
power~\cite{MembraneParadigm}
\labeq{LBZ}{
L_{\rm BZ} \approx 10^{51} \bigg(\frac{\chi}{0.85}\bigg)^2 \bigg(\frac{M_{\rm BH}}{5.6M_\odot}\bigg)^2\bigg(\frac{B_{\rm BH}}{10^{15}G}\bigg)^2 \rm erg/s,
}
where $B_{\rm BH}$ is the characteristic value of the magnetic field
on the BH horizon, and the other parameters are normalized to values
from the simulations of~\cite{prs15}. 

Jet launching requires a near force-free environment, and how close to
force-free an environment is, is indicated by the magnetization
parameter $B^2/8\pi\rho c^2$. In a sense the magnetization plays the
role of the jet trigger. Consistent with this concept, the jet found
in~\cite{prs15} was launched around the time where $B^2/8\pi\rho c^2
\gtrsim 10$. The characteristic Lorentz factors found in the expanding,
collimated outflow at the end these simulations were only mildly
relativistic ($\Gamma_L \sim 1.3$). However, when the incipient jet
was fully developed its magnetization was $B^2/8\pi\rho c^2 \sim
100$. This was a very important finding, because terminal Lorentz
factor of a magnetically-powered, axisymmetric jet is set by and
approximately equals this
number~\cite{B2_over_2RHO_yields_target_Lorentz_factor}. Hence, in
principle, the incipient jets found in~\cite{prs15} can be accelerated
to Lorentz factors required to explain sGRB phenomenology. Finally,
these calculations furnished the first computation of the delay time
between the GW peak amplitude and the EM signal.

While these results were obtained assuming initially strong magnetic
fields, they were still dynamically unimportant ($\beta \geq 20$ in
the NS interior initially), thus the outcome should be independent of
the initial magnetic field strength, because, as ~\cite{prs15} argued,
initially weak magnetic fields inside the disk should be amplified to
values $\sim 10^{15}$ G due to MRI. This expectation is confirmed by
the simulations of~\cite{kskstw15} who find that characteristic values
of the plasma parameter $\beta$ in the disk are $10^2-10^3$ (top right
panel of Fig.~\ref{fig:ksks}), and since characteristic values for the
gas pressure in these disks are $P_{\rm gas}\sim \rho v^2 \sim
10^{30}\rm dyn/cm^2 (\rho/10^{10}{\rm g/cm^3})(v/0.2c)^2$, the
expression $\beta=8\pi P_{\rm gas}/B_{\rm disk}^2$ yields for the
magnetic field ($B_{\rm disk}$) in the disk
\begin{equation}
\label{Bdisk}
B_{\rm disk} \sim 10^{15}
  \bigg(\frac{\beta}{100}\bigg)^{1/2}\bigg(\frac{P_{\rm gas}}{10^{30}
    \rm dyn/cm^2}\bigg)^{1/2} \rm G.
\end{equation} 
{\it Thus, numerical relativity BHNS simulations demonstrate that BHNS
  mergers are viable progenitors of sGRBs.}

\section{Simulations of NSNS mergers}
\label{NSNS}

It has been 17 years since successful hydrodynamics simulations of
NSNS mergers in full general relativity were carried
out~\cite{2000PhRvD..61f4001S}. Like simulations of BHNSs, simulations
of NSNS mergers in which the final outcome is a BH, involve the
modeling of physical obstacles like MHD shocks and spacetime
singularities, too.

An interesting aspect about NSNS mergers is that the outcome has more
options than BHNS mergers. The uncertainties in the nuclear EOS and
the fact that current observations allow a large range of nuclear EOSs
with Tolman-Oppenheimer-Volkoff (TOV) limit mass $\sim
2.0M_\odot-2.8M_\odot$~\cite{Lattimer2012ARNPS..62..485L,Ozel2016ARA&A..54..401O},
then depending on the binary total mass the outcome of an NSNS merger
could be any one of the following:

\begin{itemize}

\item[I)] If the binary total mass $M_{\rm NSNS}$ is less than the
  Tolman-Oppenheimer-Volkoff (TOV) limit mass, a massive spinning
  neutron star that never collapses to a BH will form.

\item[II)] If $M_{\rm NSNS}$ (minus any mass lost to escaping
  material) is greater than the TOV limit, but less than the maximum
  mass when allowing for maximal uniform rotation -- the supramassive
  limit (which is typically 20\% larger than the TOV
  limit~\cite{Morrison2004}) -- a supramassive NS (SMNS) will form. An
  SMNS is initially differentially rotating, but due to braking of the
  differential rotation by magnetic fields (or by
  viscosity)~\cite{Shapiro:2000zh,Duez2004PhRvD..69j4030D}, it will
  eventually be brought to uniform rotation, and can collapse to a BH
  (delayed collapse) following spin-down, e.g., due to magnetic dipole
  radiation.

\item[III)] If $M_{\rm NSNS}$ (minus any mass lost to escaping
  material) exceeds the supramassive limit, straight after the merger
  either a BH will form or an HMNS. A threshold mass $M_{\rm thres}$
  separates these two possibilities:

  \begin{itemize}
    \item[a)] If $M_{\rm NSNS} > M_{\rm thres}$, then a BH will form
      following merger on a dynamical time scale (prompt collapse).

    \item[b)] If $M_{\rm NSNS} < M_{\rm thres}$, then a hot,
      differentially rotating, dynamically stable HMNS will form. The
      HMNS is transient because a combination of gravitational wave
      emission, braking of the differential rotation and neutrino
      cooling will drive its eventual collapse to a BH (delayed
      collapse).

  \end{itemize}

For work on determining $M_{\rm thres}$
see~\cite{2003PhRvD..68h4020S,2006PhRvD..73f4027S,Bauswein2013PhRvL.111m1101B}.

\end{itemize}

As far as sGRBs are concerned, at this time only outcomes II) and III)
seem to be relevant, because these are the only ones where a BH-disk
engine could potentially emerge. However, numerical relativity
simulations have shown that no disk remains outside the BH which forms
following the collapse of supramassive neutron
stars~\cite{Shibata2000PhRvD..61d4012S,Shibata2003ApJ...595..992S,Baiotti2005PhRvD..71b4035B,Baiotti2007CQGra..24S.187B}
(see also ~\cite{Margalit2015PhRvL.115q1101M}). This is because the
equatorial radius of the SMNS is inside the radius which becomes the
innermost stable orbit of the remnant black hole. Thus, at this time,
it seems that only type III) outcomes can form a BH-disk engine.

Many years of numerical relativity simulations of binary NSNSs have
allowed us to gain a better understanding of the parameter space that
leads to these different outcomes. For a comprehensive review of
simulations of NSNS binaries we refer the reader
to~\cite{faber_review} (see
also~\cite{Duez2010CQGra..27k4002D,2016arXiv160703540B}). Next we will
focus primarily on work in full GR related to NSNS mergers as sGRBs.

\subsection{Hydrodynamic simulations}

Like simulations of BHNSs, the first studies of NSNS-mergers as sGRB
engines in full GR focused on cases where a BH-disk system forms after
merger, and investigated the potential for an appreciable accretion
disk to form outside the remnant black hole. The basic results in this
regard have not changed much over the last 10 years, and were
summarized in the original studies of irrotational NSNS in full GR
in~\cite{2003PhRvD..68h4020S} and~\cite{2006PhRvD..73f4027S}. There
different EOSs and mass ratios were considered, finding that
asymmetric NSNS binaries typically have larger disk masses than equal
mass binaries, and that disk masses up to $0.06M_\odot$ are possible
for mass ratios of 0.75. The authors also derived the following
EOS-dependent, and total-mass-dependent fitting formula for predicting
the disk mass following BH formation~\cite{2006PhRvD..73f4027S}
\labeq{fitNSNS}{
M_{\rm disk} = M_{\rm disk,0} + A(1-q)^p,
}
where $q \leq 1$ is the binary mass ratio, $M_{\rm disk,0}$ is the
disk mass for $q=1$, and $A,\ p$ fit parameters. For the APR
EOS~\cite{APREOS} with total mass $2.96M_\odot$, the authors find
$M_{\rm disk,0}=4\times 10^{-4}M_\odot$, $A\approx 1.44M_\odot$,
$p=4$. For the SLy EOS~\cite{SLyEOS} with total mass $2.76M_\odot$,
the authors find $M_{\rm disk,0}=3\times 10^{-4}M_\odot$, $A\approx
3.33M_\odot$, $p=3$. The range of applicability of Eq.~\ref{fitNSNS}
is restricted to $q \gtrsim 0.8$~\cite{2006PhRvD..73f4027S}.  More
recent work on this
topic~\cite{Liu2008PhRvD..78b4012L,2008PhRvD..78h4033B} is in
agreement with the earlier findings, and a different fitting formula
for disk mass predictions following the merger of quasi-circular,
irrotational (initially $\Gamma=2$ polytropic) NSNSs was derived
in~\cite{2010CQGra..27k4105R}.
%% \labeq{MdiskNSNSBaio}{ M_{\rm disk}(q,M_{\rm NSNS})=c_1(1-q)(M_{\rm
%% max}-M_{\rm NSNS})+ c_2(M_{\rm max}-M_{\rm NSNS}) } % where $M_{\rm
%% max}$ is the maximum mass of the binary, which is EOS
%% dependent. Evolving NSNS binaries with a $\Gamma=2$ $\Gamma$-law
%% EOS the authors found that $c_1=1.115 \pm 1.090$, $c_2 = 0.039 \pm
%% 0.023$. Also, for this EOS the authors find $M_{\rm
%% max}=c_3(1+q)M_{\rm TOV}$, with $c_3=1.139 \pm 0.149$. However, we
%% note that $c_3$ should be dependent on the orbital separation, too,
%% because at infinite separation $M_{\rm max}=c_3(1+q)M_{\rm TOV}$
%% While the fit of~\cite{2010CQGra..27k4105R} appears to be
%% equation-of-state independent, its applicability may be limited
%% because it was derived for a single EOS, and it is not clear that such
%% a universal fit exists. Nevertheless, in
Disk masses of up to $0.2M_\odot$ were found following asymmetric NSNS
mergers in~\cite{2010CQGra..27k4105R}. But, it is currently not known
whether such high disk masses are extreme or not. Also, the time at
which the disk mass measurement is made is usually somewhat
arbitrary. More simulations using different EOSs are necessary to draw
definitive conclusions and settle the aforementioned issues. In
addition, the impact of the NS spin on the amount of mass left to form
a disk onto the BH has not been considered yet, and it is conceivable
that spin can make a difference at least when near the threshold mass
$M_{\rm thres}$. However, simulations in full general relativity
accounting for the NS spin are still in their infancy (see
e.g.~\cite{Tichy:2011gw,Tichy:2012rp,Tsatsin2013,Bernuzzi:2013rza,PEPS2015,Kastaun2015,2015arXiv150707100D,Tacik2015,EPPS2016,EPP2016}). Finally,
simulations of eccentric NSNS that form BHs following merger, indicate
that disk masses can be up to $\sim
0.27M_\odot$~\cite{gold,East2012NSNS}. Thus, numerical relativity
simulations have established that NSNS mergers, too, can form BH-disk
engines. But, can jets emerge following an NSNS merger?

\begin{figure}
  \centering
  \includegraphics[width=0.325\textwidth]{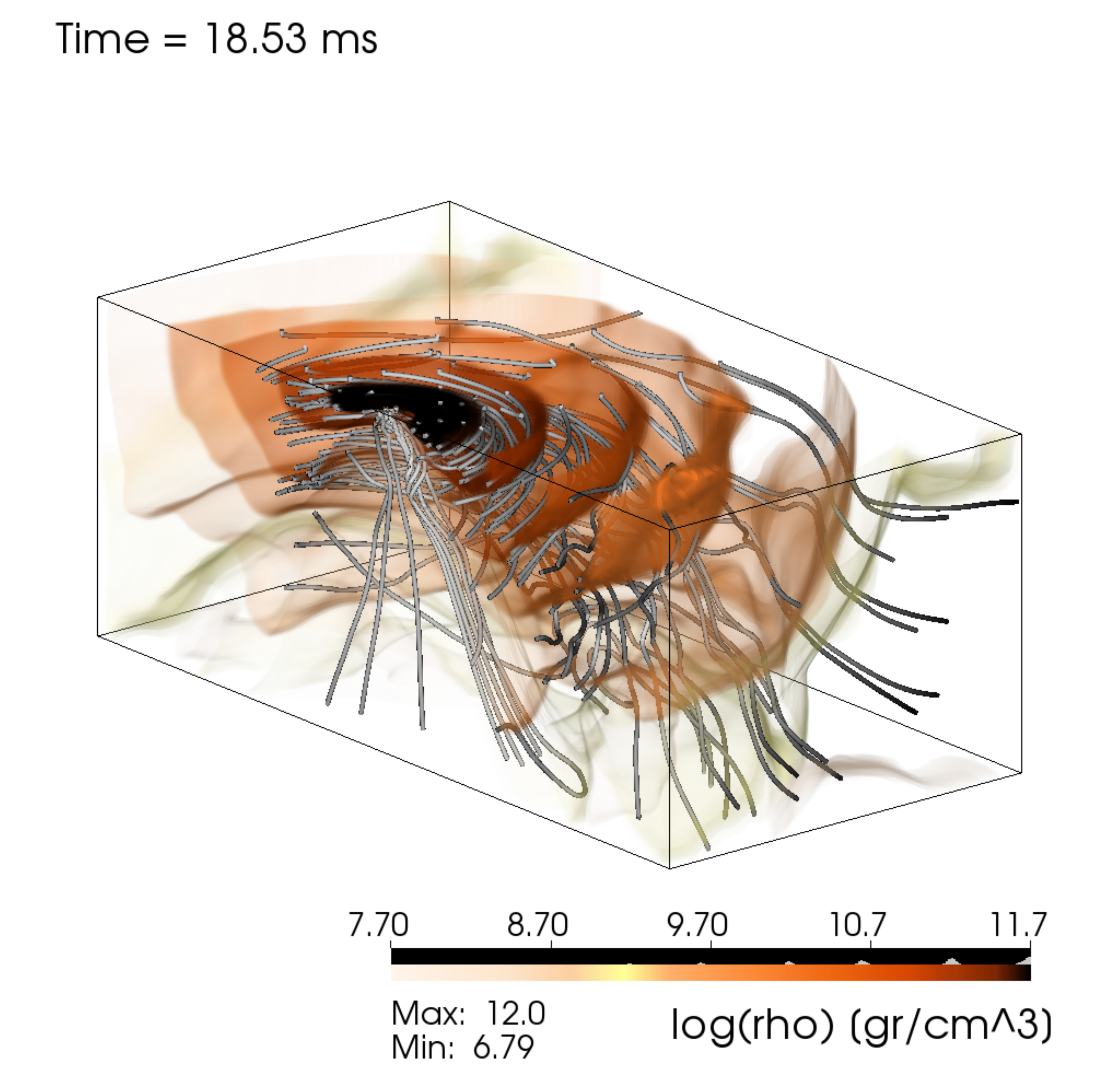}
  \includegraphics[width=0.325\textwidth]{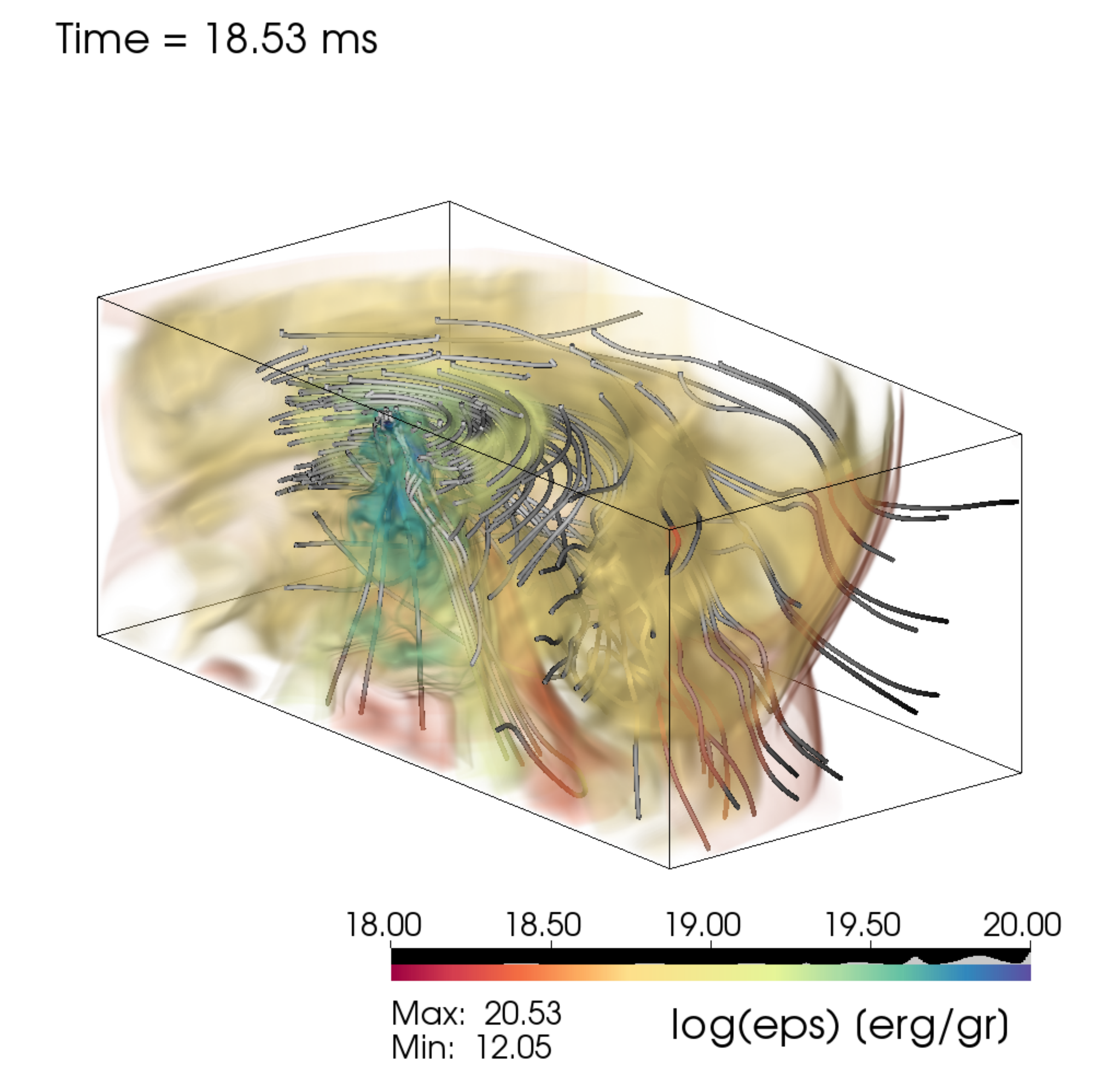}
  \includegraphics[width=0.325\textwidth]{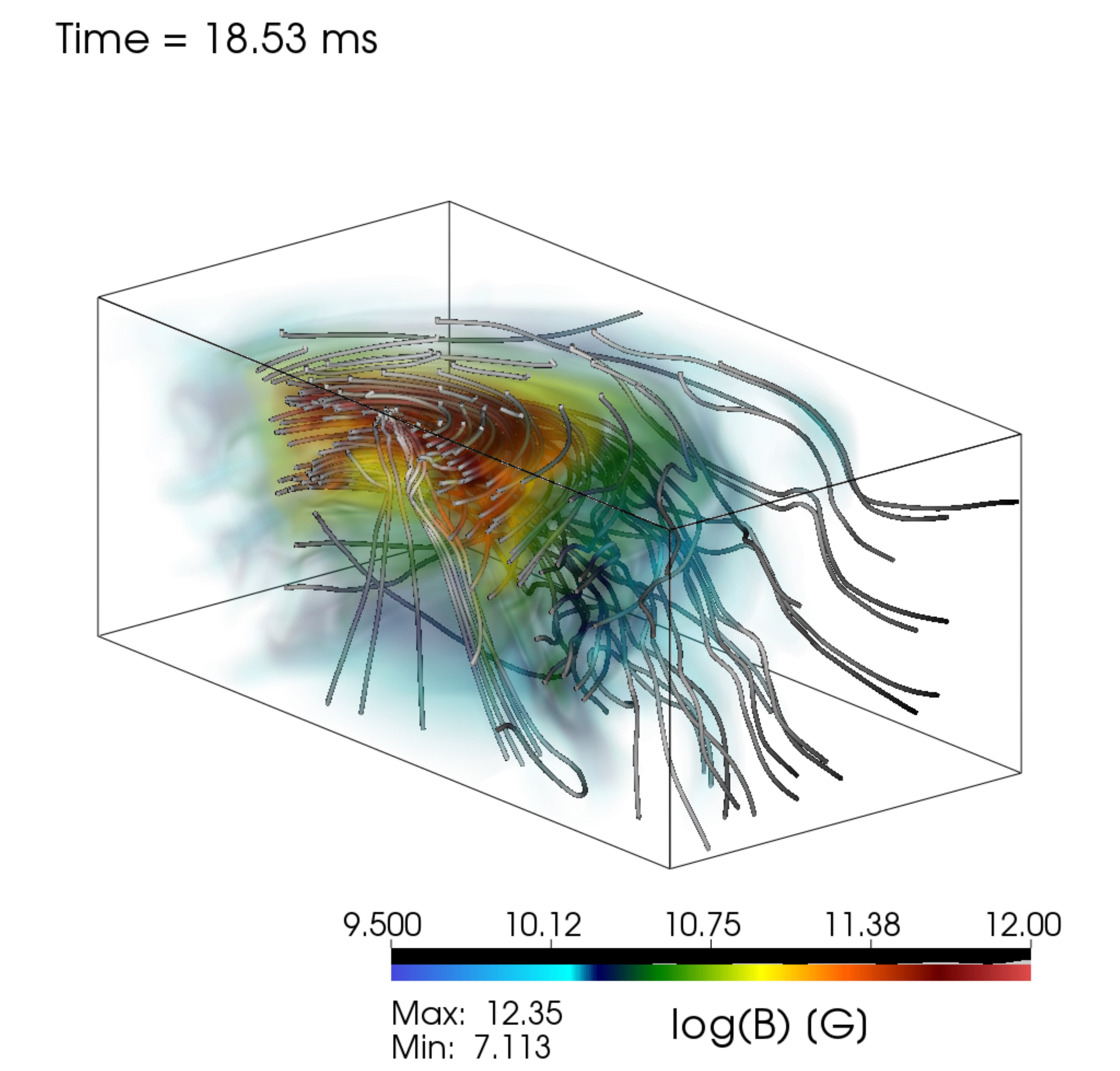}
  \caption{Volume rendering of the rest-mass density (left panel),
    specific internal energy (middle panel), and magnetic-field
    magnitude (right panel) of the BH-disk system at $t =
    18.3$ ms. The curves indicate the magnetic-field lines. Note that
    the positive z-axis points downward, and the quadrant shown has
    dimensions [0 km, 115.8 km]$\times$[-115.8 km, 115.8 km]$\times$[0
      km, 92.16 km] . Figure 7 from~\cite{Dionysopoulou2015}.
  \label{fig:Dio2015}}
\end{figure}
\begin{figure}
  \centering
  \includegraphics[width=0.325\textwidth]{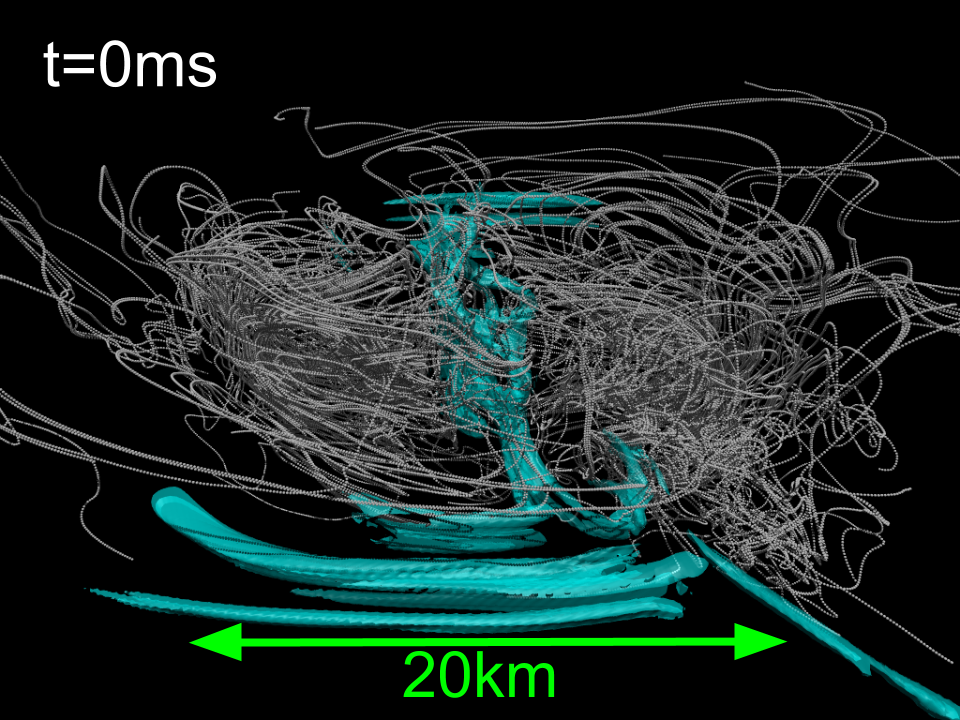}
  \includegraphics[width=0.325\textwidth]{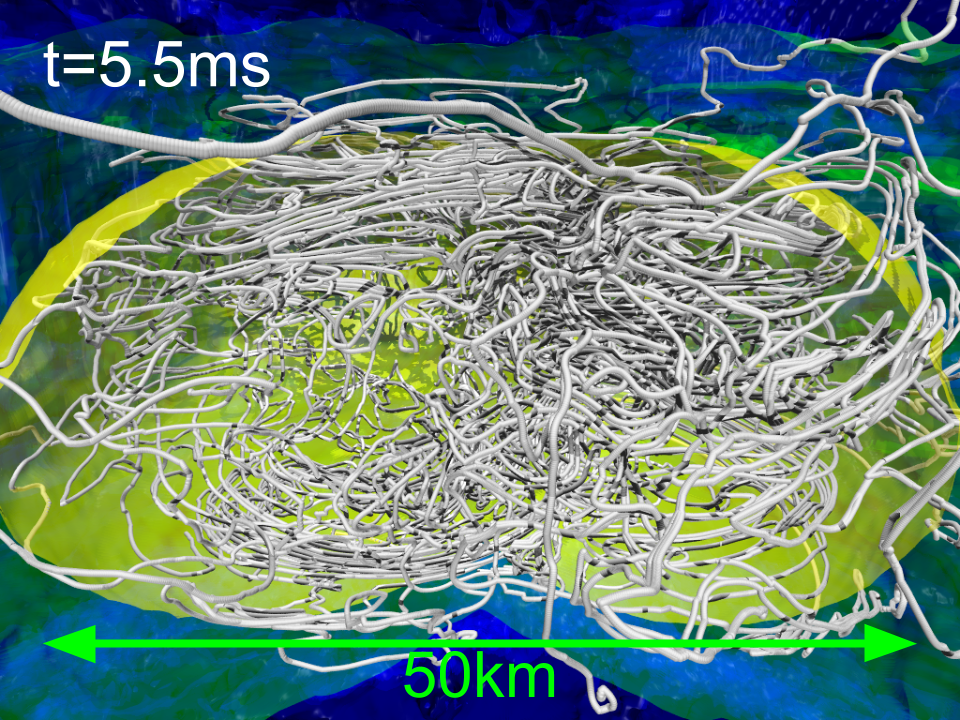}
  \includegraphics[width=0.325\textwidth]{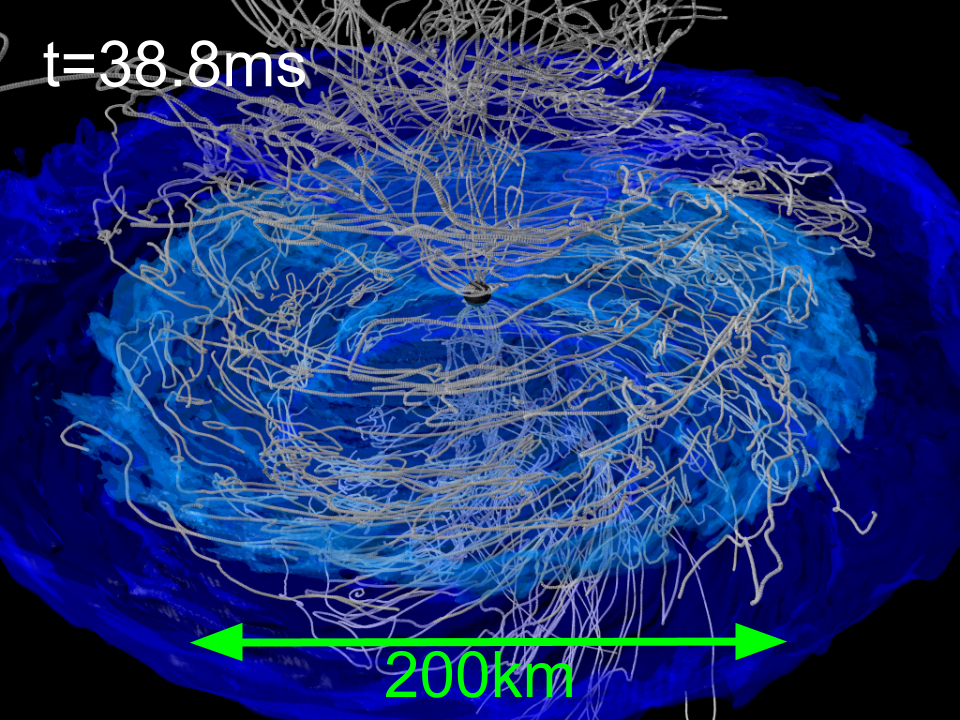}
  \caption{Volume rendering of rest-mass density, magnetic-field
    strength and magnetic-field lines (white curves) at select times.
    In the left panel, the cyan color indicates magnetic-field
    strengths greater than $10^{15.6}$ G. In the middle panel, the
    yellow, green, and dark blue colors indicate rest-mass densities of
    $10^{14}, \ 10^{12},\ 10^{10}\ g/cm^3$, respectively. In the right
    panel, the light and dark blue indicates rest-mass densities of
    $10^{10.5}$ and $10^{10}\rm g/cm^3$, respectively. Figure 1
    from~\cite{Kiuchi:2014hja}.
  \label{fig:Kiuchi2014}}
\end{figure}

Recent hydrodynamical studies of accretion onto a single BH treating
neutrinos argue that neutrino annihilation may not suffice to launch
jets following a NSNS merger. The reason is that NSNS mergers tend to
create very baryon-loaded environments~\cite{jojb15}. These results
are in agreement with the analysis
of~\cite{Murguia2014ApJ...788L...8M} who find that the post-merger
fall-back material can ``choke'' a BH-disk jet engine.
In~\cite{jojb15} it was also argued that while the environments around
BHNS mergers are not as baryon-rich, and collimated outflows can be
launched via neutrino processes, they concluded that neutrino
annihilation is an inefficient jet acceleration
mechanism. Hence,~\cite{jojb15} concluded that if jets do emerge
following compact binary mergers, then MHD processes should play a
major role in driving them.

However, the magnetic field must be able to overcome the inertia of
the matter in order to launch a jet, and achieving magnetic-field
dominance in the dense environment surrounding a NSNS merger remnant
is not trivial. One can estimate how strong the magnetic field near
the BH has to be, by equating the magnetic energy density with the
rest-mass energy density of the merger remnant
atmosphere. Relativistic NSNS simulations demonstrate that
characteristic rest-mass densities around the remnant are $\rho_0 \sim
10^9\rm g/cm^3$. Thus, the relation $B^2/8\pi \gtrsim \rho_0 c^2$
yields
\begin{equation}
\label{Bthres}
 B \gtrsim 10^{15}\bigg(\frac{\rho_0}{10^9 \rm
    g/cm^3}\bigg)^{1/2} \rm G.  
\end{equation}
But, how can a typical pulsar magnetic field be amplified from an
initial value of $10^{10}-10^{12}$ G on the NS surface to $10^{15}$ G?
In the next session we address this question and discuss whether
BH-disk engines formed in NSNS mergers can launch jets through
magnetic processes.

\subsection{Magnetohydrodynamic simulations}

\begin{figure*}
\centering
\includegraphics[width=0.4\textwidth]{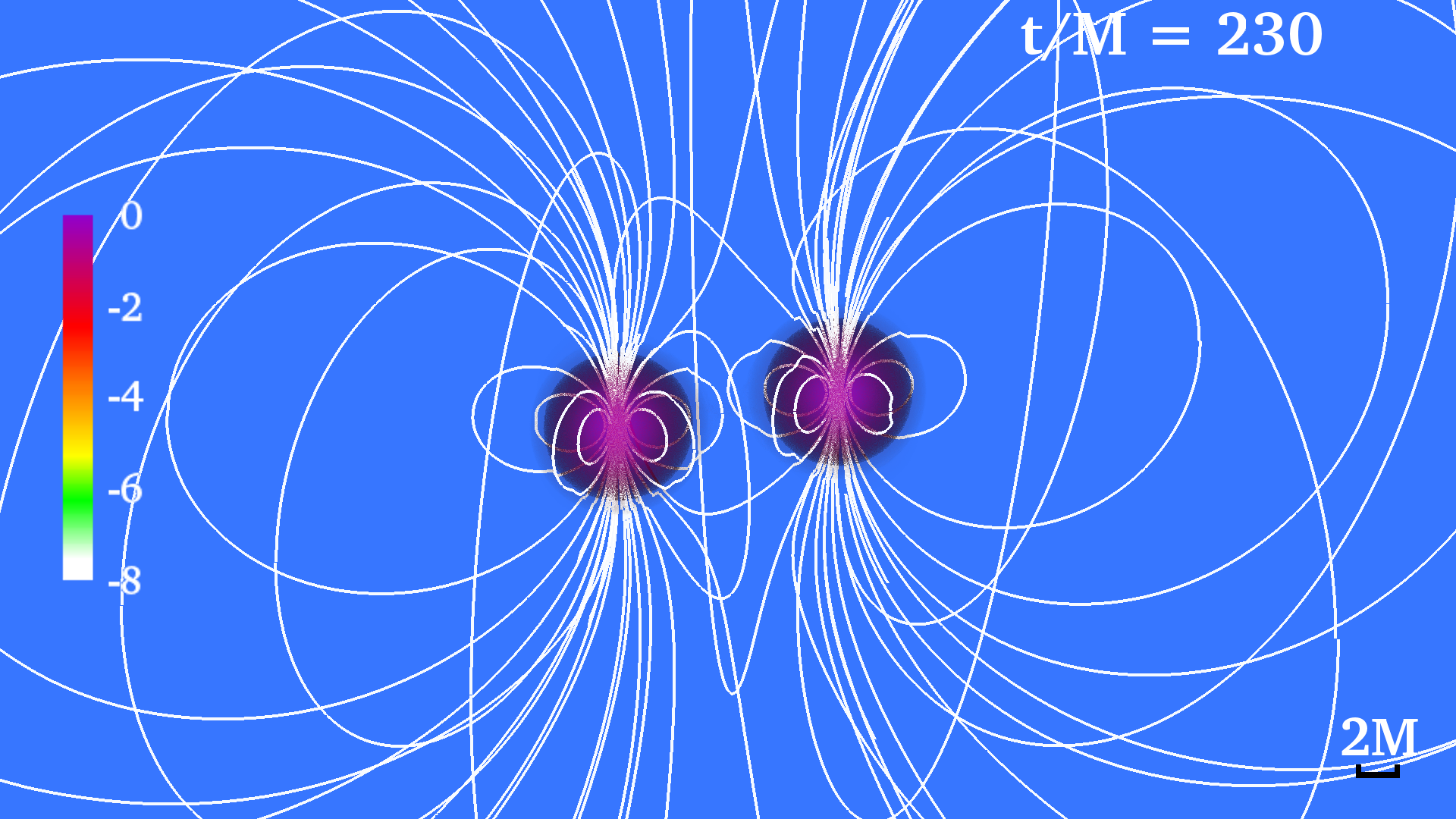}
\includegraphics[width=0.4\textwidth]{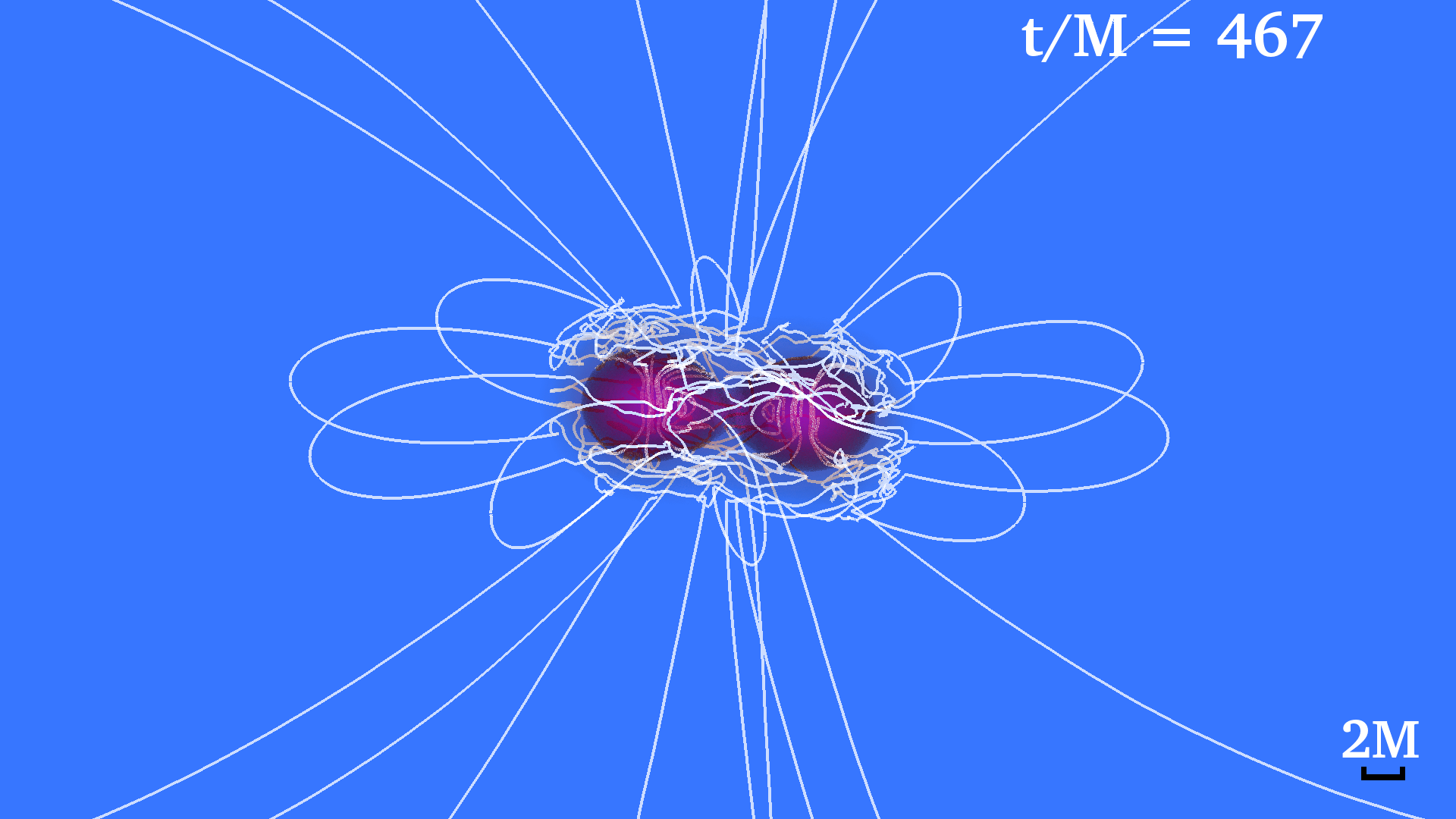}
\includegraphics[width=0.4\textwidth]{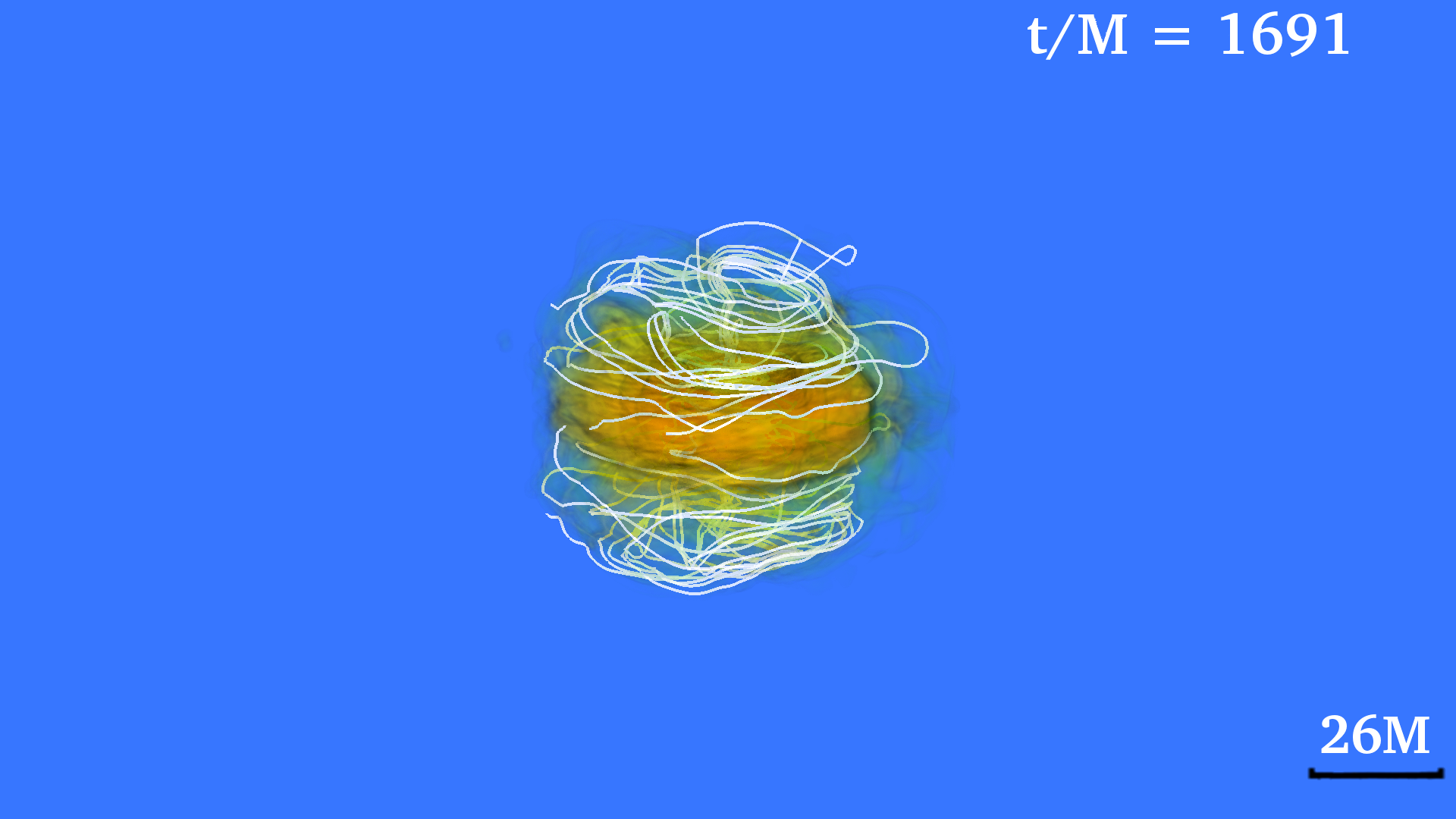}
\includegraphics[width=0.4\textwidth]{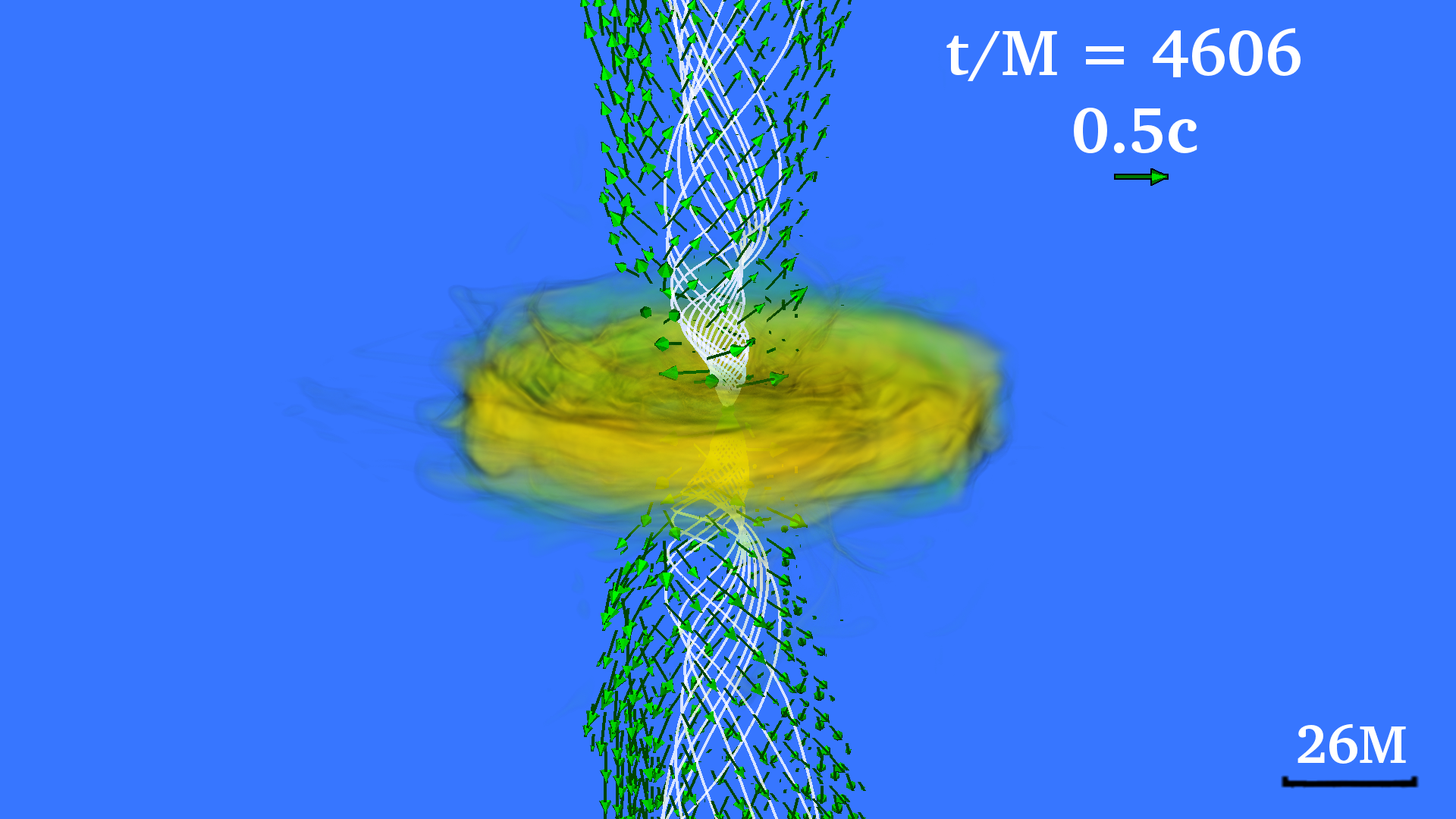}
\caption{ Same as in Fig.~\ref{fig:Binit}, but here
  $\rho_{0,{\rm max}}= 5.9\times 10^{14}(1.625\,M_\odot/M_{\rm
    NS})^2\ {\rm g cm}^{-3}$ and $M=1.47\times 10^{-2}(M_{\rm
    NS}/1.625M_\odot)$ ms = $4.43(M_{\rm NS}/1.625M_\odot)$ km. The
  appearance of an organized, large scale magnetic field inside the
  incipient jet is clear in the bottom right panel. Figure 1 from
  ~\cite{RLPSNSNSjet2016}.
\label{fig:NSNSsnapshots}}
\end{figure*}

The Newtonian simulations of~\cite{Rosswog2002MNRAS.334..481R}
reported that the Kelvin-Helmholtz instability (KHI) naturally
occurs in the shearing layer at the collision interface during an NSNS
merger. The KHI development has also been confirmed in NSNS
simulations in full
GR~\cite{ahllmnpt08,Rezzolla:2010fd}. Following~\cite{Rosswog2002MNRAS.334..481R},
the Newtonian simulations of~\cite{Price05052006} discovered that {\it
  the development of the KHI at merger can generate magnetar-level
  magnetic-field strengths within 1 ms.} Local special relativistic
ideal MHD simulations have confirmed this
picture~\cite{Zrake2013ApJ...769L..29Z}, and some works have adopted
subgrid models to simulate this effect in global NSNS simulations in
full GR~\cite{Giacomazzo2015ApJ,Palenzuela2015PhRvD..92d4045P}.
Self-consistent simulations in full GR of this effect were carried out
in~\cite{Kiuchi:2015sga}, where unprecedentedly high resolution was
adopted and was found that the KHI and the MRI occurring during and
shortly after merger amplify the magnetic fields in the HMNS to rms
values of $10^{15.5}$ G within $\sim 5$ ms.  Thus, it is fairly
established that at least for equal mass NSNS mergers the combination
of KHI and MRI are the principal hydromagnetic processes through which
the magnetic field can grow to values capable of launching jet
outflows even before the star may collapse and form a BH-disk engine.

Early long term ideal MHD simulations of NSNS mergers in full
GR~\cite{ahllmnpt08,Liu2008PhRvD..78b4012L} reported no jet
launching. On the other hand, a subsequent ideal MHD calculation of
binary NSs in full GR reported the formation of ``jet-like
structures''~\cite{ML2011}. This means the formation of a funnel-like
structure, but not the emergence of a collimated, Poynting-dominated
outflow. The same result was also found in recent resistive MHD
simulations of NSNS mergers in full GR~\cite{Dionysopoulou2015} (see
Fig.~\ref{fig:Dio2015}). A more recent full GR ideal MHD study of
NSNSs~\cite{Kiuchi:2014hja} did not find a jet or an ordered poloidal
field (see Fig.~\ref{fig:Kiuchi2014}), and concluded that the ram
pressure of the fall-back material is so strong that, in contrast to
BHNSs~\cite{kskstw15}, not even a wind can be launched after BH-disk
formation. More recent ideal MHD NSNS simulations in full
GR~\cite{2016arXiv160403445E} do not report jets or the appearance of
a large scale, ordered magnetic field following merger.  However, the
initial magnetic field strengths are low and the adopted resolution is
not high enough to capture the magnetic field growth due to KHI and
MRI.

Motivated by the successful jet launching in the BHNS calculations
of~\cite{prs15},~\cite{RLPSNSNSjet2016} adopted similar methods as
in~\cite{prs15}, but this time in a NSNS setting, focusing on the same
binary configuration as the one evolved in~\cite{ML2011}. About $\sim
60$ ms following merger, incipient jets emerge even in this NSNS
scenario (see Fig.~\ref{fig:NSNSsnapshots}). The authors also
performed a comparison study with an identical case where the initial
magnetic fields were confined in the interiors of the stars. The study
showed that jets are launched for interior only magnetic fields, too,
and on the same time scale. Consistent with the findings
of~\cite{Kiuchi:2014hja} the authors found that a jet is launched only
after the density of the fall-back matter above the BH has decreased to
levels where $B^2/(8\pi \rho c^2) \gg 1$. The disk lifetime in the
simulations of~\cite{RLPSNSNSjet2016} was estimated to be $\sim 0.2$
s, and the jet Poynting luminosity $10^{51}$ erg/s, which are again
consistent with typical sGRBs. The magnetization in the incipient jet
outflow found in the simulations was $B^2/8\pi\rho c^2 \sim 100$,
which implies that the terminal Lorentz factor of these jets can reach
$\Gamma_L ~\sim 100$ to explain sGRB phenomenology. The success of
launching jets both with interior only and interior/exterior magnetic
fields was attributed to the fact that the magnetic fields in the
scenario where an HMNS forms, can be amplified to magnetar levels
before collapse to a BH takes place.

Eq.~\eqref{LBZ} for a $10^{16}$G magnetic field on a $2.8M_\odot$ BH
with spin $\chi=0.7$ -- the values found in~\cite{RLPSNSNSjet2016} --
predicts a BZ power of
\labeq{LBZ2}{
L_{\rm BZ} \approx 10^{52} \bigg(\frac{\chi}{0.7}\bigg)^2 \bigg(\frac{M_{\rm BH}}{2.8M_\odot}\bigg)^2\bigg(\frac{B_{BH}}{10^{16}G}\bigg)^2 \rm erg/s.
}
Thus, the electromagnetic luminosity found in the simulations is close
but does not match the BZ power. Nevertheless, this mismatch could be
due to insufficient spatial resolution or the approximate nature of
Eq.~\eqref{LBZ2}, or the more baryon loaded environments surrounding
NSNS merger remnants where Eq.~\eqref{LBZ2} may not be applicable.

\begin{figure*}
\centering
\includegraphics[width=0.7\textwidth]{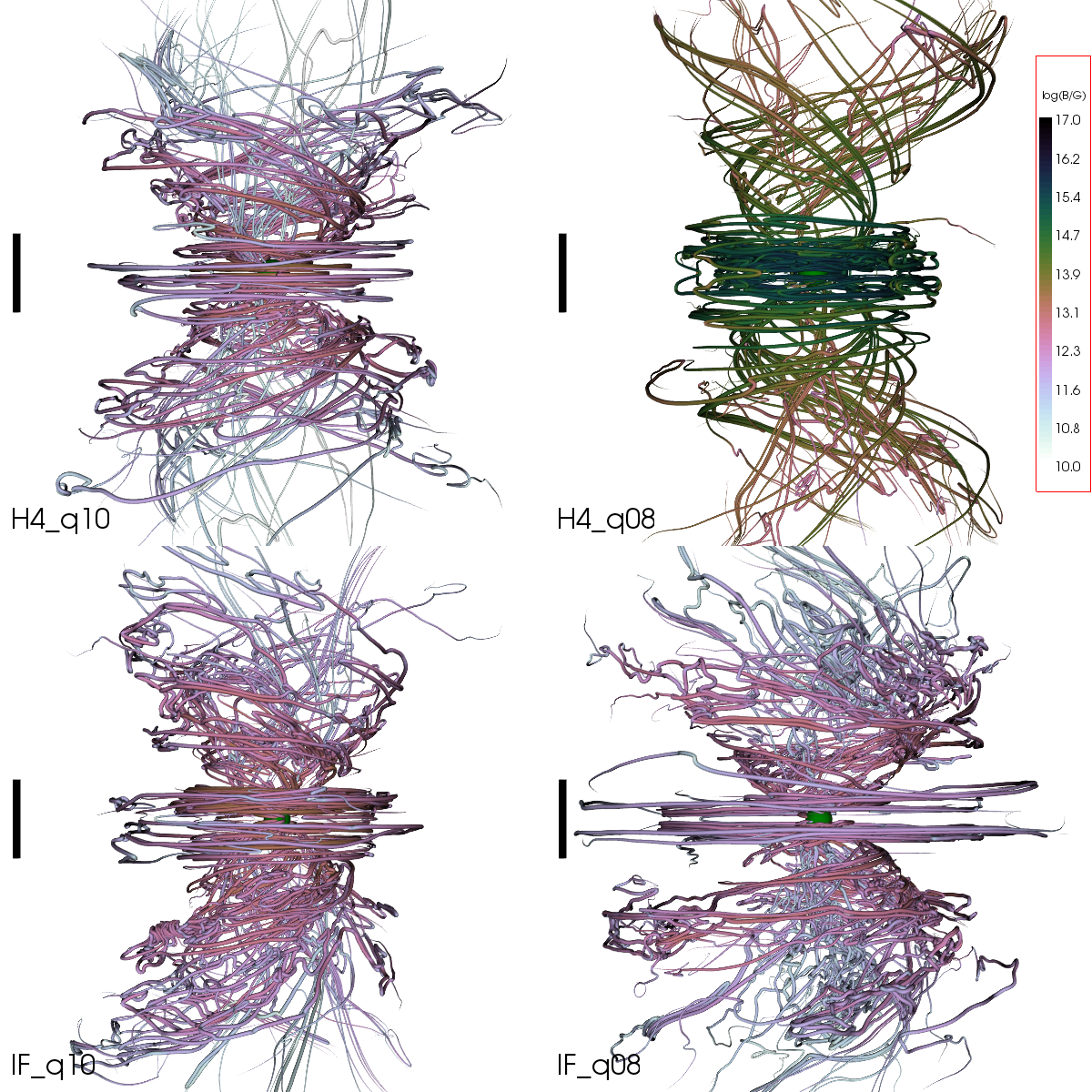}
\caption{Magnetic field lines $\sim 32$ ms following merger for
  different initial NSNS models. H4 and IF stand for different
  EOSs studied in, and ``q10'' (``q08'') indicate a mass
  ratio of 1.0 (0.8). The black vertical lines indicate a length scale
  of 20 km. The color represents the magnetic field strength
  ($\log_{10} (|B [G]|$)). Figure 16 from~\cite{2016PhRvD..94f4012K}.
\label{fig:2016PhRvD}}
\end{figure*}

We note, here, that unlike the other MHD studies of NSNS mergers in
full GR, the authors of~\cite{RLPSNSNSjet2016} seeded the initial
neutron stars with dipole magnetic fields which are high in comparison
with typical inferred magnetic field strengths of pulsars in NSNS
binaries. However strong, they were still dynamically unimportant
initially, and offered a natural means for generating
equipartition-level magnetic fields in the post-merger remnant, as is
anticipated for dynamically stable HMNSs because of the KHI and MRI
effects. Thus, what~\cite{RLPSNSNSjet2016} argued is that if the KHI
and MRI have enough time to amplify the magnetic fields prior to
collapse, then an incipient jet can emerge following collapse to BH.

A more recent work~\cite{2016PhRvD..94f4012K} performed a large suite
of magnetized NSNS simulations in full GR with magnetic fields
restricted in the stellar interiors initially, and varying the
orientation of the magnetic dipole moment (aligned/antialigned with
orbital angular momentum), the EOS and the mass ratio. The authors
report that their results confirm the emergence of an ordered magnetic
field (see Fig.~\ref{fig:2016PhRvD}), but that longer evolutions are
required for jet emergence.

Finally, it is important to clarify that there are two effects at play
following collapse of a dynamically stable HMNS, but secularly
unstable, that in a sense are competing from the point of view jet
launching. One effect is that the density in the funnel decreases with
time as the fall-back material within the funnel is accreted onto the
BH. The other effect is that at the same time the disk is being
accreted onto the BH. If the disk is accreted before the density of
the fall-back material in the funnel has decreased sufficiently, then
an sGRB jet will not be possible.  Thus, a successful magnetically
powered jet requires that the density in the funnel drop well below a
critical value (at which $B^2/8\pi\rho_{\rm crit} c^2 = 1$) on a time
scale that is shorter than the disk accretion time scale. Note that
different EOSs change not only the disk mass, but also the amount of
ejected and fall-back matter. Thus, for a given amount of disk matter
generated after NSNS mergers with different EOSs, a magnetically
powered jet may not be possible for any EOS because of potentially
different amount of fall-back material. However, neutrino annihilation
should help in lowering the density in the funnel. Thus, a combination
of neutrino and magnetic processes could potentially launch jets in
cases where the fall-back material time scale is long.

\section{Conclusions and future challenges}
\label{other}

Over the last 10 years numerical relativity simulations of BHNS and
NSNS mergers have significantly augmented our understanding of how
these compact binaries may form BH-disk engines. It is now
well-established that BH-disk systems are a generic outcome of compact
binary mergers involving neutron stars. However, forming a BH with an
accretion disk is only a necessary requirement to explain sGRBs in the
model of a hyperaccreting BH model which drives twin jets that expand
at highly relativistic velocities (a leading model for sGRBs). A
second necessary condition to establishing theoretically that BHNS and
NSNS are viable progenitors of sGRBs (in the hyperaccreting BH model)
is to show that the BH-disks their mergers give birth to can launch
jets.  Recent simulations in full GR have allowed us to study the
impact of magnetic fields and assess whether jets can be launched from
these engines. The general consensus is that following merger a large
scale ordered magnetic field can emerge that is in principle able to
drive a collimated, magnetically dominated outflow -- a jet. So far,
the only works demonstrating self-consistent jet launching following
merger have been presented in~\cite{prs15} for BHNS mergers and
in~\cite{RLPSNSNSjet2016} for NSNS mergers. These simulations
integrated for much longer times than other numerical simulations
where magnetic fields are amplified to magnetar-level strengths
following merger. At this time, it appears that BH-disk systems formed
following a BHNS merger can launch magnetically powered jets only if
the initial NS is endowed with a magnetic field that extends from the
interior out to the exterior. On the other hand, BH-disk systems
formed following a NSNS merger can launch magnetically powered jets
when the initial NS is endowed either with interior only magnetic
fields or with magnetic fields extending from the interior out to the
exterior. However, the choice of surface magnetic-field strength
in~\cite{prs15,RLPSNSNSjet2016} was larger by a couple of orders of
magnitude when compared to typical pulsar magnetic fields. This choice
was justified based on the post-merger expectations for
magnetic-field amplification by the Kelvin-Helmholtz and
magnetorotational instabilities. While the jet emergence
in~\cite{prs15,RLPSNSNSjet2016} should as a result be independent of
the initial magnetic-field strength (because their strong fields were
still dynamically unimportant initially), these calculations must be
revisited with weaker initial magnetic fields. Doing so would require
much higher resolution (almost 10 times higher) to be able to properly
capture all hydromagnetic processes. Adopting such high resolution
would render these simulations impractical with current methods and
computer resources. However, as methods become more accurate, codes
are developed to scale better, and with computer technology advances,
long-term high-resolution simulations should be within reach within
the next decade. Moreover, it is very challenging to evolve accurately
magnetic fields in low density environments with ideal MHD
codes. Thus, the calculations adopting magnetic fields that extend
from the NS interior all the way out to the NS exterior should also be
revisited with more sophisticated methods. Maybe a resistive MHD
approach would be suitable, but such schemes in full GR have been
developed~\cite{Palenzuela:2012my,Dionysopoulou2013} only recently,
and studies of BH-disk jet engines for sGRBs with resistive MHD are
completely in their infancy. However, given that from a theoretical
standpoint force-free electrodynamics is a subset of ideal
MHD~\cite{2006MNRAS.367.1797M,Paschalidis:2013gma}, in principle, one
should be able to develop an ideal MHD algorithm that can evolve
accurately both magnetically-dominated and matter-dominated
environments at the same time.

At this point, it is important to note that several sGRBs have richer
phenomenology than just the gamma-ray burst. For example, about 1/3 of
the sGRBs demonstrate strong ``afterglow'' activity for an extended
time~\cite{Gehrels2004ApJ...611.1005G}. A complete theoretical model
should be able to explain the full range of phenomenological features
sGRBs have, and perhaps explaining the burst is the easy part. To this
extend, simulations of compact binaries in full GR have revealed that
these systems exhibit richer phenomenology than just launching a
burst~\cite{Lehner:2011aa,Paschalidis:2013jsa,PalenzuelaLehner2013,Palenzuela:2013kra,Ponce:2014sza}
having both ``precursor'' and ``aftermath'' EM signals. Thus,
simulations are already providing opportunities to think about sGRBs
in a different way than the ``standard'' paradigms. Nevertheless, the
sGRB phenomenology remains poorly understood, and if we understand it,
it could provide discriminating power to choose among the different
models. It could also be that different sGRBs have different
progenitor systems.

Despite the tremendous developments in the study of compact binary
mergers as sGRB engines, many open questions still remain and
represent challenges for the next generation of compact binary
simulations in full GR. Here we give an incomplete list of such
questions: Are the incipient jets found so far stable and do they
persist for an accretion time?  What mechanism powered these
magnetically dominated jets? The BZ effect is a likely candidate, but
results so far, while strongly suggestive~\cite{prs15}, are not
conclusive. How are these incipient jets accelerated to $\Gamma_L
\gtrsim 100$? How do jets shine in gamma rays? Is the internal shock
mechanism~\cite{Meszaros:2006rc,LeeRamirezRuiz2007} realized? What is
the role of neutrinos? Can the neutrino effective bulk viscosity and
drag quench the magnetic-field growth due to MRI in a HMNS (see
e.g.~\cite{Guilet2016arXiv161008532G} and references therein)? Can
compact binary mergers account for the subclass of sGRBs with extended
X-ray emission (see
e.g.~\cite{Gehrels2004ApJ...611.1005G,SGRB_local,Rowlinson2013MNRAS.430.1061R,Gompertz2014MNRAS.438..240G})
or are other sGRB models necessary (see
e.g.~\cite{MacFadyen2005astro.ph.10192M,Metzger2008MNRAS.385.1455M,Bucciantini2012MNRAS.419.1537B,Ciolfi2015ApJ...798L..36C,Rezzolla2015ApJ...802...95R}
and references therein)? What can we infer from the time delay between
an observed GW signal and EM signal?  What is the correct, hot,
nuclear EOS? How can we combine GW and EM observations to better infer
the correct EOS? 
%Are compact binary mergers the sites where r-process elements form?

As a side discussion, we recall that there are other progenitor models
of sGRBs that have been proposed and have not received much attention
in the numerical relativity community, yet. For example, the merger of
a BH with a white dwarf (WD) was proposed
in~\cite{Janka1999ApJ...527L..39J}, as was the accretion-induced
collapse of WDs in~\cite{Vietri1999ApJ...527L..43V}. Moreover, we note
that sGRBs could be powered following massive white dwarf -- neutron
star (WDNS) mergers, if the remnant collapses to form a BH. On the
other hand, if a massive neutron star is the merger outcome, then it
may power a gamma-ray
flash~\cite{Ruffini2016arXiv160202732R}. Preliminary studies in full
GR suggest that the NS-disk remnant of a WDNS merger is supported
against collapse through a combination of additional thermal pressure,
due to shock heating, and centrifugal support from the rapid
differential rotation~\cite{2011PhRvD..84j4032P}. However, following
cooling and angular momentum redistribution the remnant can collapse
to form a BH-accretion disk
system~\cite{2009PhRvD..80b4006P,2011PhRvD..83f4002P} that may power a
gamma-ray burst. If this scenario is realized, the GRB would not take
place shortly after merger as is expected in a BHNS or NSNS merger,
but on the much longer cooling and angular momentum redistribution
time scales. Nevertheless, it is not clear yet whether this possibility
can materialize because, at least for intermediate mass white dwarfs,
nuclear burning after merger is anticipated unbind some fraction of
the WD debris (see e.g.
~\cite{Metzger2012MNRAS.419..827M,Fernandez2013ApJ...763..108F,Margalit2016MNRAS.461.1154M}
and references therein).

Finally, if the recent tantalizing Fermi detection of a hard X-ray
signal 0.4 seconds after the merger of the binary black hole event
GW150914~\cite{2016ApJ...826L...6C}, which was consistent with sky
location of GW150914, was not a chance coincidence, then it would
suggest that stellar-mass binary black hole mergers also could take
place in a circumbinary magnetized disk. Preliminary GRMHD studies of
accretion disks onto equal-mass binary black
holes~\cite{Farris:2012ux,Gold:2013zma,Gold2014} predict that $\sim
1000M$ following merger there is a boost in the Poynting luminosity of
the jet outflows observed from these systems. Interestingly, for
$M=65M_\odot$ --the inferred total mass of GW150914-- the time scale
to the luminosity boost is $1000M \sim 0.3$ s, i.e., {\it very close
  to the delay time between the GW150914 peak amplitude and the Fermi
  signal.}  Future observations will show whether the recent Fermi
detection was a chance coincidence, and if not, whether accreting BHBH
systems can explain such short-EM-burst-like events.

\ack

I am grateful to Luis Lehner and Stuart L. Shapiro for carefully
reading the manuscript and for helpful comments and suggestions. This
work was supported by NSF grant PHY-1607449, the Simons Foundation,
and NASA grant NNX16AR67G (Fermi).

\section*{References}
\bibliographystyle{plain}
\bibliography{main}

\end{document}